\documentclass[7pt, review]{elsarticle}
\usepackage{setspace}
\usepackage{enumerate}
\usepackage{graphicx}
\usepackage{amsmath, bm}
\usepackage{float}
\usepackage{mdframed}
\usepackage{multirow}
\usepackage{textcomp}
\usepackage{subcaption}
\usepackage{url}
\usepackage{makecell}
\usepackage{amssymb}
\usepackage{bm}
\usepackage{textcmds}
 \usepackage{hyperref}
\biboptions{numbers,sort&compress}

\usepackage{tikz}
\usetikzlibrary{shapes.geometric, arrows}
\usepackage{caption}  

\def\be{\begin{equation}}\def\ee{\end{equation}}
\def\ba{\begin{array}}\def\ea{\end{array}}
\def\bfg{\begin{figure}}\def\efg{\end{figure}}
\makeatletter
\def\fps@figure{htbp}
\makeatother

\usepackage{stackengine}
\stackMath
\newcommand\tenq[2][1]{%
 \def\useanchorwidth{T}%
  \ifnum#1>1%
    \stackunder[0pt]{\tenq[\numexpr#1-1\relax]{#2}}{\scriptscriptstyle\sim}%
  \else%
    \stackunder[1pt]{#2}{\scriptscriptstyle\sim}%
  \fi%
}

\RequirePackage[margin=20mm]{geometry}

\journal{}
\begin{document}

\begin{frontmatter}

\title{Wiener–Hopf Analysis of Transient Mode-I/Mode-II Fracture in Rotating Heterogeneous Magnetoelastic Orthotropic Media under Initial Stresses}

\author[label1]{Diksha}
\author[label1]{Soniya Chaudhary*}
\author[label1]{Pawan Kumar Sharma}
\cortext[cor1]{Corresponding author: soniyachaudhary18@gmail.com}
\address[label1]{Department of Mathematics and Scientific Computing, National Institute of Technology Hamirpur, Himachal Pradesh, 177005, India}

\begin{abstract}
This study investigates the fracture behavior under both opening (Mode I) and in-plane sliding (Mode II) conditions for a semi-infinite crack in a rotating, spatially graded magnetoelastic orthotropic strip with combined horizontal and vertical initial normal stresses. The strip is assumed infinite in extent, with the crack aligned along its longitudinal axis and parallel to the rotation axis. The governing field equations are reduced to an analytically tractable form by employing the Fourier transform in the spatial domain and the Laplace transform in the temporal domain. The effect of a sudden application of traction, represented by a Heaviside step function, is analyzed for both normal and shear loading conditions. The resulting boundary-value problem is addressed through the iener–Hopf method, yielding analytical representations of the stress intensity factors (SIFs). The near-tip asymptotic stress fields are evaluated to derive explicit SIF representations for each loading configuration. Laplace inversion is performed using the generalized Chebyshev–Laguerre polynomial method. Numerical simulations are conducted for $\alpha$-Uranium and Graphite--Epoxy, with isotropic materials considered for comparison to assess the role of anisotropy. 
The temporal evolution of the SIFs is investigated for varying material gradation parameters, rotation rates, magnetoelastic coupling, and initial stresses. Results reveal that functional grading significantly influences both the magnitude and time evolution of the SIFs, rotation introduces a stabilizing effect, and magnetoelasticity exhibits mode-dependent impacts. These findings provide valuable insight into fracture behavior in advanced magnetoelastic composites, with relevance to high-performance rotating structures such as surface acoustic wave devices, aircraft wings, and helicopter blades.
\end{abstract}


\end{frontmatter}
\section{Introduction}
\subsection{Spatially varying materials and their relevance to crack mechanics}
Functionally graded materials (FGMs) are advanced composites in which material properties vary gradually across the structure, typically achieved by spatially grading the volume fractions of metals and ceramics. This tailored gradation allows FGMs to simultaneously exploit the high corrosion and thermal resistance of ceramics and the mechanical toughness of metals, while reducing thermal and residual stresses~\cite{erdogan1995fracture,gu1997cracks,bahr2003cracks}. Such characteristics have driven their application in high-temperature and mechanically demanding environments, including aerospace, nuclear power, and automotive systems.
The brittle nature of ceramic phases makes FGMs susceptible to crack initiation during manufacturing or in service. Extensive studies have addressed fracture in FGMs, from foundational analytical models~\cite{erdogan1995fracture} to investigations of crack geometry effects~\cite{gu1997cracks,bahr2003cracks},  crack detection strategies~\cite{yu2009identification}, and three-dimensional analyses~\cite{zhang20113d}. These works collectively highlight the importance of accurate fracture modeling for the reliable and durable design of graded structures in critical applications.
While much of the early fracture mechanics research focused on isotropic approximations, the inherent anisotropy in certain FGMs—especially those with directional microstructural gradations—necessitates a deeper understanding of orthotropic and anisotropic fracture behavior.

\subsection{Fracture mechanics in orthotropic and anisotropic media}
Orthotr- opic materials constitute a special class of anisotropic solids in which mechanical and thermal properties vary independently along three mutually orthogonal axes of material symmetry~\cite{sadowski2009cracks,singh2025review}. This structural anisotropy originates from microstructural arrangements such as fiber reinforcement, layered laminates, or naturally occurring crystalline formations. In special orthotropy, these principal material directions coincide with orthogonal planes of symmetry, which dictate the elastic and fracture response.
High-performance orthotropic composites, such as prepregs, carbon-fiber laminates, and epoxy-based systems, are increasingly employed in aerospace components, corrosion-resistant marine structures, load-bearing automotive parts, and tooling for metal and polymer forming~\cite{gay2022composite,parveez2022scientific}. Their anisotropic stiffness and strength also make them suitable analogues for geological formations, where rocks and crystalline minerals form composite-like structures. In seismology, the Earth's crust and near-surface layers can be modeled as anisotropic media, with earthquake rupture processes exhibiting fracture propagation in Mode~II, Mode~III, or mixed Mode~II–III~\cite{aspri2022dislocations,reches2023earthquakes}. The semi-infinite crack serves as a canonical model for these rupture phenomena and has further applications in structural engineering, seismic hazard mitigation, and geophysical modeling~\cite{rubio2000dynamic,karan2024interaction}.
Fracture mechanics in orthotropic media has been investigated for several decades, beginning with the multiple-crack interaction analysis in anisotropic plates~\cite{sadowski2009cracks}. Thermoelastic effects in orthotropic fracture were addressed using coupled-field formulations~\cite{zhong2012thermoelastic}, while idealized crack problems in layered orthotropic media were analyzed through Fourier–Laplace transform techniques~\cite{ayhan2006fracture}. Subsequent work examined fiber–crack orientation influences on fracture parameters~\cite{fakoor2017influence} and mixed-mode propagation with T-stress effects in orthotropic plates~\cite{farid2019mixed}. 
These studies collectively demonstrate that anisotropic fracture behavior cannot be deduced from isotropic theory through simple parameter substitution. Instead, the coupled influence of material symmetry, loading mode, and crack geometry must be explicitly incorporated in predictive models. 

\subsection{Semi-infinite crack and mixed mode loading}
The theoretical insights gained from orthotropic fracture mechanics directly inform the analysis of more specific geometries such as the semi-infinite crack, which serves as a canonical model for understanding localized fracture processes in anisotropic and graded media~\cite{xu2008dynamic,ustinov2020orthotropic}. A semi-infinite crack is idealized as extending indefinitely in one direction while remaining bounded in the other, making it a valid approximation for long flaws where boundary effects are negligible~\cite{knauss1966stresses,rubio2000dynamic}.
Such configurations are encountered in practice in diverse engineering settings: fatigue cracks in aerospace wing skins~\cite{rubio2000dynamic}, long weld defects in ship hull plates~\cite{das2024study}, delamination fronts in laminated composites~\cite{ustinov2020orthotropic}, and interface cracks in thermal barrier coatings on turbine blades~\cite{ustinov2019semi}. In many of these cases, mixed-mode loading arises naturally from asymmetric boundary conditions, combined thermal–mechanical effects, or non-uniform far-field stresses~\cite{xu2008dynamic,farid2019mixed}. The simultaneous presence of Mode~I and Mode~II loading significantly alters stress intensity factors and crack growth stability~\cite{rubio2000dynamic,chen2014propagation}. Analytical approaches—particularly Fourier–Laplace transforms and Wiener–Hopf factorization—have proven indispensable for deriving benchmark solutions that guide numerical simulations and aid in designing fracture-resistant composite and graded structures~\cite{krylov1977handbook,ustinov2019semi}.
The analytical study of semi-infinite cracks has evolved over several decades. Knauss~\cite{knauss1966stresses} provided one of the earliest treatments, analyzing stresses in an infinite medium with a semi-infinite fracture under static loading. Rubio-Gonzalez and Mason~\cite{rubio2000dynamic} extended the analysis to transient loading in orthotropic materials, highlighting the significant impact of anisotropy on transient fracture behavior.
 Xu \textit{et al.}~\cite{xu2008dynamic} extended this analysis to orthotropic functionally graded materials, showing how gradation alters dynamic crack-tip fields.

Subsequent work broadened the scope to coupled-field and interface fracture problems. Chen \textit{et al.}~\cite{chen2014propagation} investigated Mode~I propagation of conducting semi-infinite cracks in piezoelectric media, capturing electromechanical coupling effects. Ustinov~\cite{ustinov2019semi} derived closed-form solutions for semi-infinite interface fracture in bi-material elastic layers, while Ustinov \textit{et al.}~\cite{ustinov2020orthotropic} examined central semi-infinite cracks in orthotropic strips under arbitrary remote loading. More recent studies include Singh \textit{et al.}~\cite{singh2020semi}, who analyzed moving semi-infinite cracks in orthotropic strips, and Das and Tanwar~\cite{das2024study}, who addressed interfacial semi-infinite cracks in composite structures under mixed-mode conditions. Das \textit{et al.}~\cite{das2024wiener} applied the Wiener–Hopf technique to solve the anti-plane problem for a semi-infinite fracture in orthotropic composites, providing closed-form analytical solutions.
Karan \textit{et al.}~\cite{karan2024interaction} investigated the interaction of SH-waves with a semi-infinite crack in anisotropic media, revealing how wave–crack coupling affects dynamic fracture behavior. Alam \textit{et al.}~\cite{alam2024magnetoelastic} extended this framework to include magnetoelastic effects, studying the influence of anti-plane waves on crack propagation in orthotropic structures under coupled mechanical–magnetic fields.

\subsection{Instantaneous application of traction in dynamic fracture problems}
The sudden application of traction to a crack surface represents a critical transient loading scenario in dynamic fracture mechanics, where rapid energy release excites high-frequency stress waves whose interaction with material anisotropy and crack geometry governs the early fracture response \cite{ing2001transient,flitman1963waves,khimin2024analysis,bresciani2025quasistatic}. 
Such loading conditions occur in diverse high-consequence applications, including low-velocity impact on aerospace composite skins during maintenance operations~\cite{abrate1991impact,shi2017modelling}, impulsive hydrodynamic pressures acting on ship hull weld defects following underwater explosions, thermal shock-induced delamination in turbine blade coatings~\cite{suresh1998fatigue,he2025damage}, and seismic rupture initiation along fault segments, modeled as semi-infinite cracks subjected to shear traction release~\cite{karan2024interaction}. 
Theoretically, the problem is posed as a mixed initial–boundary value problem, with the instantaneous load specifying the initial condition and the crack faces, either traction-free or displacement-prescribed, defining the boundary conditions \cite{wang2001dynamic}.
Analytical techniques—most notably the Fourier–Laplace transforms combined with Wiener–Hopf factorization—yield analytical benchmark solutions for semi-infinite cracks in isotropic and anisotropic solids~\cite{knauss1966stresses,rubio2000dynamic,xu2008dynamic, das2024wiener}, revealing that stress intensity factors under sudden traction exhibit strong temporal oscillations, mode coupling, and sensitivity to wave speeds and crack orientation. In orthotropic laminates, the directional stiffness mismatch leads to mode conversion between longitudinal and shear waves, modifying near-tip stress amplitudes and potentially amplifying certain fracture modes~\cite{fakoor2017influence}. For multifunctional materials such as piezoelectrics or magnetoelastic composites, the applied traction further induces coupled electromechanical or magnetoelastic fields~\cite{chen2014propagation,alam2024magnetoelastic}, complicating the transient response. Recent advances~\cite{das2024wiener,singh2020semi} extend the analysis to moving semi-infinite cracks under sudden traction, showing that the interplay between crack motion and transient loading can cause stress-wave focusing—a phenomenon of significance in predicting structural failure and modeling earthquake rupture dynamics.

\subsection{Research gap and objectives of the study}

Despite the significant progress in understanding dynamic fracture under sudden traction in various materials and configurations, important gaps remain—particularly regarding the simultaneous influence of material heterogeneity, rotation of the medium, and initial stresses on transient fracture behavior. While prior research has examined semi-infinite cracks in anisotropic and piezoelectric media~\cite{alam2025mode, alam2024magnetoelastic, singh2020semi}, the effects of rotation and continuous spatial variation of material properties (i.e., functional grading) have largely been overlooked. Moreover, initial normal stresses, which are often present in practical scenarios due to manufacturing or operational conditions, have received limited attention in the transient fracture context~\cite{ing2001transient, xu2008dynamic, chen2014propagation}.
To address these deficiencies, the present study focuses on the transient propagation of a semi-infinite crack in a functionally graded magnetoelastic orthotropic strip undergoing uniform rotation about its longitudinal axis, incorporating initial stresses in both horizontal and vertical directions. By employing the Wiener–Hopf technique in the Laplace domain and subsequently performing numerical inversion, explicit dynamic stress intensity factors (SIFs) for Mode-I and Mode-II loading under sudden uniform traction are derived. This formulation allows a comprehensive parametric investigation into how material heterogeneity, rotation, magnetoelastic coupling, and initial stresses collectively influence fracture dynamics.
Ultimately, the objective is to extend the existing theoretical frameworks to capture these coupled effects in more realistic and technologically relevant settings, enhancing the predictive capability for the design and analysis of advanced composite and multifunctional materials.

\subsection{Motivation}  
Modern aerospace and energy systems demand materials that can withstand extreme mechanical, thermal, and electromagnetic environments without compromising performance. In a spinning turbine rotor or helicopter blade, rotation is not just motion—it reshapes the material’s stiffness through centrifugal forces and redirects stress waves via Coriolis effects, altering fracture initiation and growth \cite{deshpande2018effect}. In the realm of magnetoelastic devices, such as surface acoustic wave (SAW) sensors and MRI-compatible actuators, the elastic–magnetic coupling becomes the heartbeat of signal accuracy and sensitivity, especially when functional grading is used to fine-tune properties across complex temperature and stress fields \cite{luo2024magnetoelectric,wolframm2022high}.  
Yet, the life of a component is rarely free from hidden stresses. Manufacturing processes, rapid thermal cycling, or years of operational loading can leave behind residual stress patterns that quietly shift stress intensity factors (SIFs) and fracture toughness \cite{jweeg2020dynamic,valverde2022influence}. These intertwined influences—rotation, functional grading, magnetic coupling, and pre-existing stress—are not abstract concepts; they are the invisible forces that determine whether a generator shaft spins flawlessly for decades or a maglev train component fails prematurely.  
A comprehensive understanding of these coupled effects underpins the design of aerospace components with enhanced fracture resistance, magnetoelastic sensors with stable accuracy under harsh operating conditions, and rotating machinery with superior reliability in demanding environments.
\subsection{Novelty and application of the present study}
The present work addresses a research gap in the analytical and numerical modeling of plane wave propagation in rotating functionally graded orthotropic magnetoelastic materials with a semi-infinite crack, subjected to initial stress. Unlike existing studies that primarily focus on isotropic or purely elastic configurations, this formulation incorporates the combined effects of \textit{functional grading}, \textit{rotation}, \textit{magnetoelastic coupling}, and \textit{pre-existing initial stress}, thereby enabling a more accurate prediction of wave–crack interaction in realistic engineering environments. The mathematical model transforms the governing equations into the Laplace domain and then uses the Wiener–Hopf technique, facilitating the determination of stress intensity factors (SIFs) under complex loading conditions.

From an application perspective, the study has direct relevance to high performance rotating systems such as aerospace turbine rotors, helicopter blades \cite{quaranta2014rotorcraft}, and MRI-compatible actuators \cite{senturk2016design}, where rotation-induced centrifugal stiffening and Coriolis effects interact with magnetic fields to influence wave dispersion and fracture characteristics. In magnetoelastic surface acoustic wave (SAW) sensors \cite{chen2007wave}, particularly those designed for biomedical diagnostics and structural health monitoring, functional grading can be tailored to enhance sensitivity and stability under varying thermal and mechanical gradients. Moreover, the results are pertinent to maglev train components \cite{xiang2022dynamic}, rotating generator shafts \cite{jweeg2020dynamic}, and other magneto-mechanical systems, where initial stress significantly modifies fracture toughness and device reliability. The integrated modeling framework proposed herein can thus aid in the optimization of material architecture and operational parameters for both mechanical durability and functional performance.

\subsection{Layout of the manuscript}
The structure of the manuscript is outlined as follows: Section~\ref{Mathematical Modelling for Plane Strain Problem} presents the formulation of the plane strain problem, accounting for magnetoelastic coupling, rotational effects, and initial stresses, where both displacement and stress fields are expressed in the Laplace domain.
Subsection~\ref{Mode I Loading} addresses Mode-I (opening mode) crack behavior and derives the stress intensity factor (SIF) using the Wiener–Hopf technique, while Section~\ref{Mode-II Loading} focuses on Mode-II (sliding mode) and provides explicit SIF expressions using the same approach.
Section~\ref{Numerical Simulations and Graphical Representation} covers the numerical simulations and graphical results. Subsection~\ref{Laplace Inversion} presents the computation of inverse Laplace transforms using series expansions in terms of generalized Chebyshev–Laguerre polynomials, implemented in MATLAB.
and Subsection~\ref{graphical} illustrates the temporal variation of the SIF under different fracture parameters.
Finally, Section~\ref{Conclusion} summarizes the main findings and key insights of the study.

\section{Mathematical modelling for plane strain problem}
\label{Mathematical Modelling for Plane Strain Problem}
\begin{figure}
    \centering
    \includegraphics[width=1\linewidth]{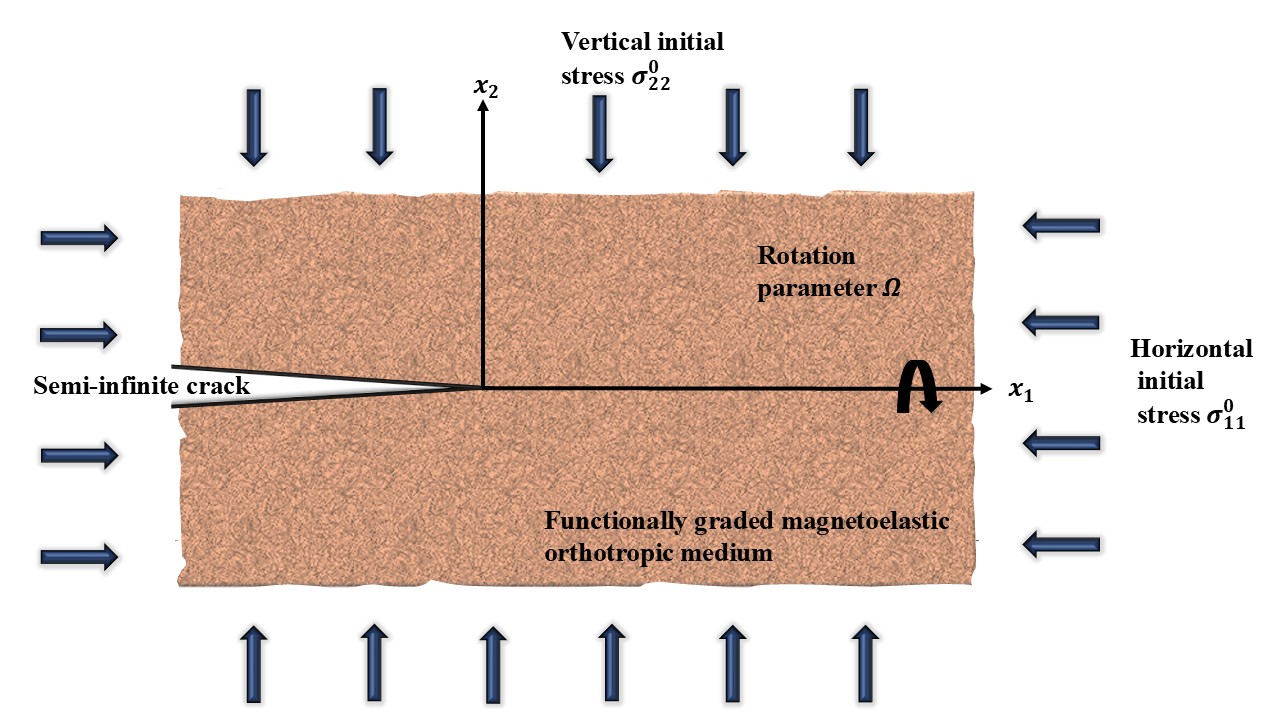}
    \caption{Schematic of a crack of semi-infinite extent in a magnetoelastic orthotropic medium with spatially varying properties. }
    \label{geometry}
\end{figure}
\bfg[htbp]
\centering
\begin{subfigure}[b] {0.47\textwidth}
\includegraphics[width=\textwidth ]{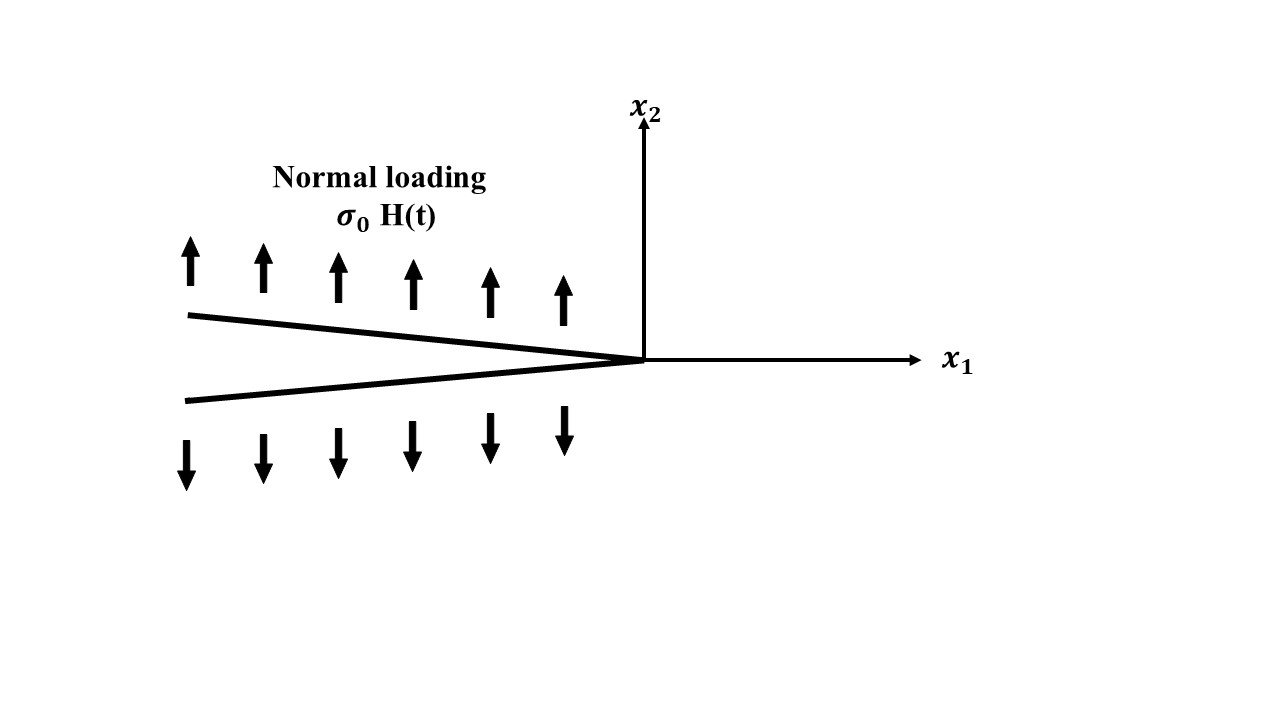}
\caption{}
\label{slide1}
\end{subfigure}
~
\begin{subfigure}[b] {0.47\textwidth}
\includegraphics[width=\textwidth ]{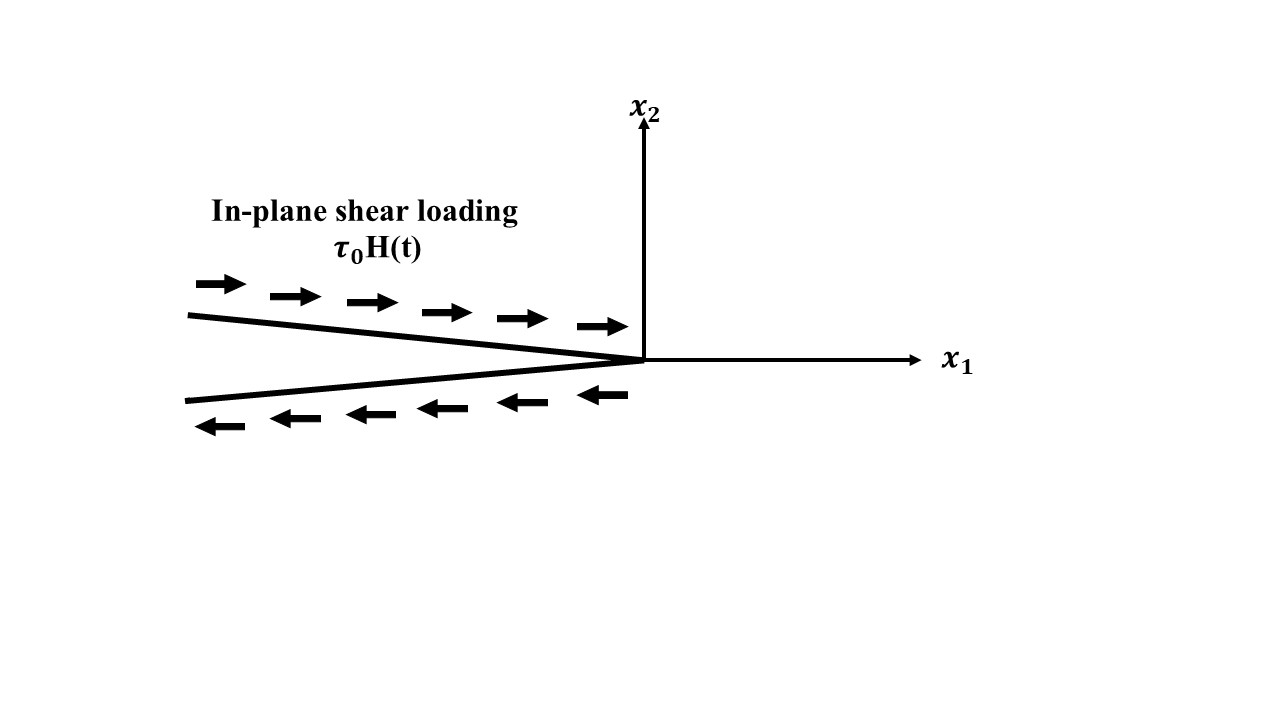}
\caption{}
\label{slide2}
\end{subfigure}

\caption{Configuration of the semi-infinite crack under  (a) opening (Mode I) and (b) sliding (Mode II) conditions.}
\label{slide12}
\efg
The problem considers a crack of semi-infinite extent in a magnetoelastic orthotropic medium, as illustrated in Fig.~\ref{geometry}.
Let \(x_i\) \((i=1,2,3)\) represent the Cartesian coordinates describing the spatial position within the medium shown in Fig.~\ref{geometry}. 
The \(x_1\)-axis is taken along the direction of the crack, with the crack tip located at \(x_1=0\), and the medium is considered unbounded in this direction. 
The \(x_2\)-axis is oriented perpendicular to the crack plane, while the \(x_3\)-axis is aligned with the out-of-plane direction, forming a right-handed coordinate system.
Deformation is restricted to the $x_1x_2$-plane, corresponding to plane strain conditions associated with Mode~I (opening) and Mode~II (sliding) fractures. Accordingly, the displacement vector $\mathbf{u} = (u_1, u_2, u_3)$ can be written in terms of its components as:
\begin{equation}
u_1 = u_1(x_1, x_2, t), \quad 
u_2 = u_2(x_1, x_2, t), \quad 
u_3 = 0, \quad 
\frac{\partial}{\partial x_3}(\cdot) = 0,
\end{equation}
indicating that displacement occurs only in \(x_1\) and \(x_2\) directions, while out-of-plane displacement and variation along the \(x_3\)-direction are neglected.

The material medium is assumed to be functionally graded, magnetoelastic, and orthotropic in nature. It is initially subjected to a uniform normal pre-stress along the $x_1$- and $x_2$-directions. In addition, the medium undergoes uniform rotation about the $x_1$-axis. The crack is considered to extend semi-infinitely in the negative $x_1$-direction  
($x_1 < 0$), thereby weakening the medium along the interface $x_2 = 0$.
The mechanical loading on the crack faces depends on the fracture mode being considered. For mode-I fracture, uniform tensile tractions are suddenly applied normal to the crack faces (see Fig. \ref{slide1}), while for mode-II fracture, uniform shear tractions are applied tangentially (see Fig. \ref{slide2}).

The behavior of the orthotropic magnetoelastic material is described by the linear constitutive relations, which can be expressed compactly in matrix form~\cite{shi2003indentation}:
\begin{equation}
\mathbf{A} = \mathbf{B} \mathbf{C},
\label{constitutive_relation_matrix}
\end{equation}
where \(\mathbf{A}\) denotes the \textit{stress vector} comprising the six independent stress components expressed in Voigt notation, and \(\mathbf{C}\) represents the corresponding \textit{strain vector}, including the six independent strain components, with the shear strains scaled by a factor of two to conform to engineering strain conventions:
\begin{align}
\mathbf{A} &= 
\begin{bmatrix}
\sigma_{11} & \sigma_{22} & \sigma_{33} & \sigma_{23} & \sigma_{13} & \sigma_{12}
\end{bmatrix}^{T},
\\
\mathbf{C} &= 
\begin{bmatrix}
\varepsilon_{11} & \varepsilon_{22} & \varepsilon_{33} & 2\varepsilon_{23} & 2\varepsilon_{13} & 2\varepsilon_{12}
\end{bmatrix}^T.
\end{align}
The matrix \(\mathbf{B}\) is the \textit{elastic stiffness matrix} that characterizes the mechanical response of the orthotropic material. It consists of stiffness coefficients \(\mu_{ij}\) arranged in a symmetric form that relates the normal and shear components of strain to the corresponding stresses:
\[
\mathbf{B} =
\begin{bmatrix}
\mu_{11} & \mu_{12} & \mu_{13} & 0 & 0 & 0 \\
\mu_{21} & \mu_{22} & \mu_{23} & 0 & 0 & 0 \\
\mu_{31} & \mu_{32} & \mu_{33} & 0 & 0 & 0 \\
0 & 0 & 0 & \mu_{44} & 0 & 0 \\
0 & 0 & 0 & 0 & \mu_{55} & 0 \\
0 & 0 & 0 & 0 & 0 & \mu_{66}
\end{bmatrix}.
\]
Here, \(\sigma_{ij}\) and \(\varepsilon_{ij}\) represent the stress and strain tensor components, respectively, and \(\mu_{ij}\) represent the elastic stiffness coefficients specific to the orthotropic material, capturing its anisotropic mechanical behavior.


The equation of motion describing small elastic disturbances in an orthotropic magnetoelastic medium, subjected to initial stresses, electromagnetic body forces \((\mathbf{J} \times \mathbf{B})\), and rotational effects characterized by the angular velocity vector \(\boldsymbol{\Omega} = (\Omega_1, \Omega_2, \Omega_3)\), is given by \cite{biot1940influence,chaudhary2025integral}:
\begin{align}
\sigma_{ij,j} + (\mathbf{J} \times \mathbf{B})_i 
&+ \frac{\partial}{\partial x_j} \Big( \bar\sigma^0_{jk} \omega_{ik} 
+ \bar\sigma^0_{ij} \varepsilon_{ij} \delta_{ij} - \bar\sigma^0_{ik} \varepsilon_{kj} \Big) \nonumber \\
&= \rho \frac{\partial^2 u_i}{\partial t^2} 
+ 2 \rho (\boldsymbol{\Omega} \times \dot{\mathbf{u}})_i 
+ \rho (\boldsymbol{\Omega} \times (\boldsymbol{\Omega} \times \mathbf{u}))_i.
\label{governing eq}
\end{align}

Here, $\rho$ denotes the mass density of the medium. 
The term $(\mathbf{J} \times \mathbf{B})i$ represents the $i$-th component of the Lorentz force, i.e., the electromagnetic body force generated due to the interaction between the induced electric current density $\mathbf{J}$ and magnetic flux density $\mathbf{B}$.
The tensor $\bar\sigma^0_{ij}$ incorporates the effects of pre-existing stresses on the incremental elastic motion.
The term \( (\boldsymbol{\Omega} \times \dot{\mathbf{u}})_i \) corresponds to the \emph{Coriolis acceleration}, which arises from the interaction between the angular velocity vector \(\boldsymbol{\Omega}\) and the particle velocity \(\dot{\mathbf{u}}\), thereby altering the trajectory of motion in the rotating frame. In contrast, the term \( (\boldsymbol{\Omega} \times (\boldsymbol{\Omega} \times \mathbf{u}))_i \) represents the \emph{centrifugal acceleration}, which acts radially outward from the axis of rotation and significantly influences the stress and displacement distribution within the medium.

The rotation and strain tensors are defined, respectively, as:
\[
\omega_{ij} = \frac{1}{2}(u_{i,j} - u_{j,i}), \quad\varepsilon_{ij} = \frac{1}{2}(u_{i,j} + u_{j,i}).
\]
Here, $(\,\cdot\,)_{,i}$ represents partial differentiation with respect to the spatial coordinate $x_i$.
Since the material is functionally graded, the properties of the material are considered to continuously vary along the length of the medium, i.e., in the $x_1$-direction. Specifically, the shear moduli $\mu_{ij}(x_1)$, the initial stress components $\bar\sigma_{ij}^0(x_1)$, and the mass density $\rho(x_1)$ are considered to follow an exponential variation along $x_1$:
\begin{equation}
\left[\mu_{ij}(x_1), \: \bar\sigma_{ij}^0(x_1),\: \rho(x_1) \right] = \left[\mu_{ij}^{'}, \: \sigma_{ij}^0,\: \rho^{'}\right] \, e^{\alpha x_1},
\label{functional grading}
\end{equation}
where $\mu_{ij}^{'}$, $\sigma_{ij}^0$, and $\rho^{'}$ are the corresponding reference values of shear modulus, initial stress, and density at $x_1 = 0$, and $\alpha$ is the material grading parameter that controls the rate of exponential variation.
The governing Maxwell’s equations for electromagnetism, are given by \cite{chaudhary2025crack}:
\begin{equation}
\begin{aligned}
\nabla \times \mathbf{E} = -\frac{\partial \mathbf{B}}{\partial t}, \quad \nabla \cdot \mathbf{B} = 0, \quad
\mathbf{B} = \mu_e \mathbf{H}, \quad
\nabla \times \mathbf{H} = \mathbf{J}, \quad
\mathbf{J} = \sigma \left( \mathbf{E} + \frac{\partial \mathbf{u}}{\partial t} \times \mathbf{B} \right),
\label{maxwell}
\end{aligned}
\end{equation}
where $\mathbf{E}$ is the electric field vector, $\mathbf{H}$ is the magnetic field intensity, and $\mu_e$ is the magnetic permeability of the material. The electrical conductivity of the medium is denoted by $\sigma$.

Assuming that the total magnetic field $\mathbf{H} = (H_{x_1}, H_{x_2}, H_{x_3})$ consists of a uniform primary field and a small perturbed component:
\begin{equation}
\mathbf{H} = \mathbf{H}^0 + \mathbf{h},
\end{equation}
where $\mathbf{H}^0 = (0, 0, H^0)$ represents the initial magnetic field aligned in the $x_3$-direction, and $\mathbf{h} = (h_{x_1}, h_{x_2}, h_{x_3})$ denotes the perturbed magnetic field due to elastic disturbances in the medium.
Now, from the Maxwell’s equations presented in Eq.~(\ref{maxwell}), and under the consideration that the magnetic field is perturbed as $\mathbf{H} = \mathbf{H}_0 + \mathbf{h}$ with $\mathbf{H}_0 = (0, 0, H_0)$, it can be deduced that the perturbed magnetic field satisfies the following equation:
\begin{equation}
\nabla^2 \mathbf{H} = \mu_e \sigma \left( \frac{\partial \mathbf{H}}{\partial t} - \nabla \times \left( \frac{\partial \mathbf{u}}{\partial t} \times \mathbf{H} \right)\right),
\label{eq5}
\end{equation}
where 
\begin{equation}
\nabla \times \left( \frac{\partial \mathbf{u}}{\partial t} \times \mathbf{H} \right) = \left(0,\, 0,\, -H^0 \left( \frac{\partial \dot{u}_{1}}{\partial x_1} + \frac{\partial \dot{u}_{2}}{\partial x_2} \right) \right),
\end{equation}
where $\dot{(\,\cdot\,)}$ denotes the time derivative. 
As the medium is assumed to be perfectly conducting, the electrical conductivity $\sigma \rightarrow \infty$. Under this condition,
\begin{align}
\frac{\partial H_{x_1}}{\partial t} = 0,  \quad
\frac{\partial H_{x_2}}{\partial t} = 0,  \quad
\frac{\partial H_{x_3}}{\partial t} = -H^0 \left( \frac{\partial \dot{u}_{1}}{\partial x_1} + \frac{\partial \dot{u}_{2}}{\partial x_2} \right). \label{HZ_eq}
\end{align}
From Eq. (\ref{HZ_eq}), it is evident that there is no perturbation in the $H_{x_1}$ and $H_{x_2}$ components of the magnetic field over time. This implies that the corresponding perturbed components $H_{x_1}$ and $H_{x_2}$ remain zero. However, the $H_{x_3}$ component evolves with time, indicating the presence of a non-zero perturbation in the $x_3$-component of the magnetic field:
\begin{align}
\frac{\partial h_{x_1}}{\partial t} = 0,  \quad
\frac{\partial h_{x_2}}{\partial t} = 0,  \quad
\frac{\partial h_{x_3}}{\partial t} = -H^0 \left( \frac{\partial \dot{u}_{1}}{\partial x_1} + \frac{\partial \dot{u}_{2}}{\partial x_2} \right). \label{hZ_eq}
\end{align}
By integrating Eq. (\ref{hZ_eq}) over time, the perturbed components of the magnetic field are obtained as:
\begin{align}
h_{x_1} = 0, \quad h_{x_2} = 0, \quad
h_{x_3} = -H^0 \left( \frac{\partial u_{1}}{\partial x_1} + \frac{\partial u_{2}}{\partial x_2} \right). \label{hz_final}
\end{align}
Therefore, under the assumption of a perfectly conducting medium and using the relationship: \[ \mathbf{J} \times \mathbf{B} = \mu_e \left[ (\nabla \times \mathbf{h}) \times \mathbf{H} \right], \] the components of the Lorentz force vector along the coordinate axes are expressed as:
\begin{align}
(\mathbf{J} \times \mathbf{B})_{x_1} &= \mu_e (H^0)^2 \left( \frac{\partial^2 u_{1}}{\partial x_1^2} + \frac{\partial^2 u_{2}}{\partial x_1 \partial x_2} \right), \\
(\mathbf{J} \times \mathbf{B})_{x_2} &= \mu_e (H^0)^2 \left( \frac{\partial^2 u_{1}}{\partial x_1 \partial x_2} + \frac{\partial^2 u_{2}}{\partial x_2^2} \right), \\
(\mathbf{J} \times \mathbf{B})_{x_3} &= 0. \label{JxB_x3}
\end{align}
Considering the initial stresses \( \bar\sigma_{11}^0 \) and \( \bar\sigma_{22}^0 \) as the normal pre-stresses acting along the \( x_1 \)- and \( x_2 \)-axis, respectively, and assuming a uniform rotation of the medium about the \( x_1 \)-axis with angular velocity vector \( \boldsymbol{\Omega} = \Omega(1, 0, 0) \).
By substituting the constitutive relations from Eq.~\eqref{constitutive_relation_matrix}, the spatially varying material properties defined in Eq.~\eqref{functional grading}, and the Lorentz force components derived in Eq.~\eqref{JxB_x3} into the general elastodynamic equation of motion presented in Eq.~\eqref{governing eq}, the governing equation for the displacement field in the magnetoelastic, pre-stressed, and rotating orthotropic medium becomes:
\begin{equation}
\mathbf{M}_{xx} \frac{\partial^2 \mathbf{u}}{\partial x_1^2}
+ \mathbf{M}_{xy} \frac{\partial^2 \mathbf{u}}{\partial x_1 \partial x_2}
+ \mathbf{M}_{yy} \frac{\partial^2 \mathbf{u}}{\partial x_2^2}
+ \alpha\, \mathbf{K}_x \frac{\partial \mathbf{u}}{\partial x_1}
+ \alpha\, \mathbf{K}_y \frac{\partial \mathbf{u}}{\partial x_2}
= \boldsymbol{\rho} \frac{\partial^2 \mathbf{u}}{\partial t^2}
+ \mathbf{F}_\Omega \mathbf{u},
\label{matrix_form_gv12}
\end{equation}
where
\[
\mathbf{u} =
\begin{bmatrix}
u_1 \\[2pt]
u_2
\end{bmatrix},
\quad
\mathbf{M}_{xx} =
\begin{bmatrix}
\mu_{11}^{'} + \mu_e (H^0)^2 + \sigma_{11}^0 & 0 \\[2pt]
0 & \mu_{66}^{'} + \sigma_{11}^0
\end{bmatrix},
\] \[
\mathbf{M}_{xy} =
\begin{bmatrix}
0 & \mu_{12}^{'} + \mu_{66}^{'} + \mu_e (H^0)^2 \\[2pt]
\mu_{12}^{'} + \mu_{66}^{'} + \mu_e (H^0)^2 & 0
\end{bmatrix},
\]
\[
\mathbf{M}_{yy} =
\begin{bmatrix}
\mu_{66}^{'} + \sigma_{22}^0 & 0 \\[2pt]
0 & \mu_{22}^{'} + \mu_e (H^0)^2 + \sigma_{22}^0
\end{bmatrix},
\]
\[
\mathbf{K}_x =
\begin{bmatrix}
\mu_{11}^{'} + \sigma_{11}^0 & 0 \\[2pt]
0 & \mu_{66}^{'} + \sigma_{11}^0
\end{bmatrix},
\quad
\mathbf{K}_y =
\begin{bmatrix}
0 & \mu_{12}^{'} \\[2pt]
\mu_{66}^{'} & 0
\end{bmatrix},
\]
\[
\boldsymbol{\rho} =
\rho^{'}\,
\begin{bmatrix}
1 & 0 \\[2pt]
0 & 1
\end{bmatrix},
\quad
\mathbf{F}_\Omega =
\begin{bmatrix}
0 & 0 \\[2pt]
0 & -\rho^{'} \Omega^2
\end{bmatrix}.
\]
In Eq. (\ref{matrix_form_gv12}), the time-dependent terms can be eliminated by applying the Laplace transform, defined as:
\begin{equation}
\mathcal{L}\{g(t)\} = \bar{g}(s) = \int_0^\infty g(t) e^{-st} \, dt, \quad
\mathcal{L}^{-1}\{\bar{g}(s)\} = g(t) = \frac{1}{2\pi i} \int_{\text{Br}} \bar{g}(s) e^{st} \, ds, \label{laplace}
\end{equation}
where \( s \) is the complex Laplace parameter, and \( \text{Br} \) denotes Bromwich contour of integration in complex \( s \)-plane.

Applying Laplace transform defined in Eq.~\eqref{laplace} to Eq. (\ref{matrix_form_gv12}), and under the assumption of zero initial conditions for velocity and displacement fields (i.e., \( u_i(x_1, x_2, 0) = 0 \) and \( \dot{u}_i(x_1, x_2, 0) = 0 \)), the governing equations are converted from time domain to Laplace domain. Denoting Laplace-transformed displacement fields by \( \bar{u}_1(x_1, x_2, s) \) and \( \bar{u}_2(x_1, x_2, s) \), the resulting system becomes:
\begin{equation}
\mathbf{\bar M}_{xx} \frac{\partial^2 \bar{\mathbf{u}}}{\partial x_1^2}
+ \mathbf{\bar M}_{xy} \frac{\partial^2 \bar{\mathbf{u}}}{\partial x_1 \partial x_2}
+ \mathbf{\bar M}_{yy} \frac{\partial^2 \bar{\mathbf{u}}}{\partial x_2^2}
+ \alpha\, \mathbf{K}_x \frac{\partial \bar{\mathbf{u}}}{\partial x_1}
+ \alpha\, \mathbf{K}_y \frac{\partial \bar{\mathbf{u}}}{\partial x_2}
= \mathbf{R}(s,\Omega)\,\bar{\mathbf{u}},
\label{laplace_matrix_form}
\end{equation}
where
\[
\bar{\mathbf{u}} =
\begin{bmatrix}
\bar{u}_1 \\[2pt]
\bar{u}_2
\end{bmatrix},
\quad
\mathbf{\bar M}_{xx} =
\begin{bmatrix}
\mu_{11}^{'} + \mu_e (H^0)^2 + \sigma_{11}^0 & 0 \\[2pt]
0 & \mu_{66}^{'} + \sigma_{11}^0
\end{bmatrix},
\]
\[
\mathbf{\bar M}_{xy} =
\begin{bmatrix}
0 & \mu_{12}^{'} + \mu_{66}^{'} + \mu_e (H^0)^2 \\[2pt]
\mu_{12}^{'} + \mu_{66}^{'} + \mu_e (H^0)^2 & 0
\end{bmatrix},
\]
\[
\mathbf{\bar M}_{yy} =
\begin{bmatrix}
\mu_{66}^{'} + \sigma_{22}^0 & 0 \\[2pt]
0 & \mu_{22}^{'} + \mu_e (H^0)^2 + \sigma_{22}^0
\end{bmatrix},
\]
\[
\mathbf{K}_x =
\begin{bmatrix}
\mu_{11}^{'} + \sigma_{11}^0 & 0 \\[2pt]
0 & \mu_{66}^{'} + \sigma_{11}^0
\end{bmatrix},
\quad
\mathbf{K}_y =
\begin{bmatrix}
0 & \mu_{12}^{'} \\[2pt]
\mu_{66}^{'} & 0
\end{bmatrix},
\]
\[
\mathbf{R}(s,\Omega) =
\begin{bmatrix}
\rho^{'} s^2 & 0 \\[2pt]
0 & \rho^{'} s^2 - \rho^{'} \Omega^2
\end{bmatrix}.
\]
To solve the system of differential equations in Eq.~\eqref{laplace_matrix_form}, a Fourier transform is applied with respect to the spatial coordinate $x_1$.
 The Fourier transform and its corresponding inverse can be expressed as follows:
\begin{equation}
\mathcal{F}\{f(x_1)\} = F(p) = \int_{-\infty}^{\infty} f(x_1) e^{i p x_1} \, dx_1, \quad
f(x_1) = \frac{1}{2\pi} \int_{-\infty}^{\infty} F(p) e^{-i p x_1} \, dp, \label{fourier_def}
\end{equation}
where $p$ denotes the Fourier transform parameter. The Laplace-domain displacement components $\bar{u}_1(x_1, x_2, s)$ and $\bar{u}_2(x_1, x_2, s)$ are then expressed as:

\be
\begin{bmatrix}
\bar{u}_1 (x_1, x_2, s) \\[2pt] \bar{u}_2 (x_1, x_2, s)
\end{bmatrix}
=
\frac{1}{2\pi}\int_{-\infty}^{\infty}
\begin{bmatrix}
e^{-ipx_1} & 0 \\
0 & e^{-ipx_1}
\end{bmatrix}
\begin{bmatrix}
M(p,x_2,s) \\[2pt] N(p,x_2,s)
\end{bmatrix}\,dp.
\label{u1u2_fourier}
\ee
where \( M(p, x_2, s) \) and \( N(p, x_2, s) \) are assumed to satisfy the subsequent ordinary differential equations:
\begin{equation}
\begin{bmatrix}
\dfrac{d^2}{dx_2^2} - B_1 & A_1 \dfrac{d}{dx_2} \\
A_2 \dfrac{d}{dx_2} & \dfrac{d^2}{dx_2^2} - B_2
\end{bmatrix}
\begin{bmatrix}
M \\[4pt]
N
\end{bmatrix}
=
\begin{bmatrix}
0 \\[4pt]
0
\end{bmatrix}
\label{ode12}
\end{equation}
where,
\begin{align*}
    A_1&=\frac{\alpha \mu_{12}^{'}-i p( \mu_{12}^{'}+\mu_{66}^{'}+\mu_e {H^0}^2)}{\mu_{66}^{'}+\sigma_{22}^0}, \: A_2= \frac{\alpha \mu_{66}^{'}-i p( \mu_{12}^{'}+\mu_{66}^{'}+\mu_e {H^0}^2)}{\mu_{22}^{'}+ \mu_e {H^0}^2+\sigma_{22}^0},\\
    B_1&=\frac{p^2( \mu_{11}^{'}+\mu_e {H^0}^2+\sigma_{11}^0)+i p \alpha(\mu_{11}^{'}+\sigma_{11}^0)+\rho^{'}s^2}{\mu_{66}^{'}+\sigma_{22}^0},\\ \: B_2&=  \frac{p^2( \mu_{66}^{'}+\sigma_{11}^0)+i p \alpha(\mu_{66}^{'}+\sigma_{11}^0)+\rho^{'}(s^2-\Omega^2)}{\mu_{22}^{'}+ \mu_e {H^0}^2+\sigma_{22}^0}.
\end{align*}
Considering that the displacements vanish as \(x_2 \to \infty\), the general solution of the ordinary differential equations \eqref{ode12} can be expressed as:
\begin{align}
\begin{bmatrix}
M(p, x_2, s) \\
N(p, x_2, s)
\end{bmatrix}
&=
\begin{bmatrix}
e^{-\beta_1 x_2} & e^{-\beta_2 x_2} \\
-\gamma_1 e^{-\beta_1 x_2} & -\gamma_2 e^{-\beta_2 x_2}
\end{bmatrix}
\begin{bmatrix}
M_1(p, s) \\
M_2(p, s)
\end{bmatrix}
\label{MNsolution}.
\end{align}
where \( M_1(p,s)\) and \( M_2(p,s)\) are arbitrary functions to be determined from boundary and continuity conditions, with \( \gamma_j(p, s) \) defined as:
\be
\gamma_j(p,s)= \frac{B_1- \beta_j^2}{A_1 \beta_j}; \quad  j=1,2 
\ee
with $\beta_1^2$  and $\beta_2^2$ being two distinct roots of the biquadratic equation:
\be
\beta^4-(A_1A_2+B_1+B_2)\beta^2+B_1B_2=0
\ee
The expressions for the displacement and stress components in the Laplace domain are obtained as:
\begin{align}
\begin{bmatrix}
\bar{u}_1(x_1, x_2, s) \\
\bar{u}_2(x_1, x_2, s)
\end{bmatrix}
&= \frac{1}{2\pi} \int_{-\infty}^{\infty}
\begin{bmatrix}
1 & 1 \\
-\gamma_1  & -\gamma_2 
\end{bmatrix}
\mathbf{M}(p,s,x_2)
 \, dp, \label{displacemnt_laplace} \\[2ex]
\begin{bmatrix}
\bar{\sigma}_{11}(x_1, x_2, s) \\
\bar{\sigma}_{22}(x_1, x_2, s) \\
\bar{\sigma}_{12}(x_1, x_2, s)
\end{bmatrix}
&= \frac{e^{\alpha x_1}}{2\pi} \int_{-\infty}^{\infty}
\mathbf{W}(p)
\mathbf{M}(p,s,x_2)
e^{-ipx_1} \, dp. \label{stress_laplace}
\end{align}
where
\[
\mathbf{W}(p) =
\begin{bmatrix}
\beta_1 \gamma_1 \mu_{12}^{'} - i p \mu_{11}^{'} &
\beta_2 \gamma_2 \mu_{12}^{'} - i p \mu_{11}^{'} \\
\beta_1 \gamma_1 \mu_{22}^{'} - i p \mu_{12}^{'} &
\beta_2 \gamma_2 \mu_{22}^{'} - i p \mu_{12}^{'} \\
\mu_{66}^{'} (i p \gamma_1 - \beta_1) &
\mu_{66}^{'} (i p \gamma_2 - \beta_2)
\end{bmatrix}, \quad
\mathbf{M}(p,s,x_2) =
\begin{bmatrix}
M_1(p,s) e^{-\beta_1 x_2} \\
M_2(p,s) e^{-\beta_2 x_2}
\end{bmatrix}.
\]
This section presents the derivation of displacement and stress fields in the Laplace domain. In the subsequent section, different modes of crack propagation, specifically Mode I and Mode II, are examined under the application of instantaneous traction on the crack faces, with respective crack loading, i.e., normal and shear, according to the propagation mode. The Wiener--Hopf technique is employed to obtain closed-form expressions for the stress intensity factors corresponding to both Mode~I and Mode~II.

\section{Mode I loading}
\label{Mode I Loading}
In the case of Mode~I (opening mode) fracture, the faces of the crack are subjected to an instantaneous application of a uniform normal traction \(\sigma_0\) at time \(t = 0\).  
Due to the geometric symmetry of the crack configuration, it is sufficient to confine the analysis to the upper half-space \((x_2 \geq 0)\).  
Accordingly, the boundary conditions prescribed along the crack plane \((x_2 = 0)\) are given by:
\begin{enumerate}
    \item \textbf{Normal stress condition on the crack faces ($x_1 < 0$):}  
    A uniform tensile traction of magnitude $\sigma_0$ is applied instantaneously at $t = 0$,  
    acting normal to the crack surfaces. This is expressed as  
    \begin{equation}
        \sigma_{22}(x_1, 0, t) = -\sigma_0 H(t), \quad x_1 < 0,
        \label{bc_modeI_normal}
    \end{equation}
    where $H(t)$ is Heaviside unit step function, indicating that load is applied suddenly  
    and maintained thereafter. The negative sign denotes that the traction is tensile in nature.

    \item \textbf{Shear stress condition along the interface ($x_1$ unrestricted):}  
    No tangential (shear) tractions are present on the crack plane, hence  
    \begin{equation}
        \sigma_{12}(x_1, 0, t) = 0, \quad \forall \quad  x_1.
        \label{bc_modeI_shear}
    \end{equation}

    \item \textbf{Displacement continuity condition ahead of the crack tip ($x_1 > 0$):}  
    The crack faces are closed ahead of the crack tip, implying zero normal separation.  
    Thus,  
    \begin{equation}
        u_2(x_1, 0, t) = 0, \quad x_1 > 0.
        \label{bc_modeI_disp}
    \end{equation}
\end{enumerate}
 Applying the Laplace transform as defined in (\ref{laplace}) to the boundary conditions (\ref{bc_modeI_normal})-(\ref{bc_modeI_disp}) results in the following expressions:
\begin{align}
\bar\sigma_{22}(x_1, 0, s) &= -\frac{\sigma_0}{s}, && \text{for } x_1 < 0, \label{lap_bc_sigma22} \\
\bar\sigma_{12}(x_1, 0, s) &= 0, && \text{for all } x_1, \label{lap_bc_sigma12} \\
\bar{u}_2(x_1, 0, s) &= 0, && \text{for } x_1 > 0. \label{lap_bc_u2}
\end{align}
Now, imposing the boundary condition given in Eq.~\eqref{lap_bc_sigma12} (i.e., \( \bar{\sigma}_{12}(x_1, 0, p) = 0 \)) into Eq.~\eqref{stress_laplace}, the following relation is obtained:
\begin{equation}
M_2(p,s)= -\eta_1 M_1(p,s), \quad \text{where} \quad\eta_1 = \frac{i p \gamma_1 - \beta_1}{i p \gamma_2 - \beta_2}. \label{M2_eta}
\end{equation}
Thus, the transformed displacement and stress components are obtained as follows:
\begin{align}
\begin{bmatrix}
\bar{u}_1(x_1, x_2, s) \\
\bar{u}_2(x_1, x_2, s) \\
\bar{\sigma}_{11}(x_1, x_2, s) \\
\bar{\sigma}_{22}(x_1, x_2, s) \\
\bar{\sigma}_{12}(x_1, x_2, s)
\end{bmatrix}
&= \frac{1}{2\pi} \int_{-\infty}^{\infty} 
\mathbf{K}(p, x_2, s) \,
M_1(p,s) \, e^{-ipx_1} \, dp,
\label{lap_Disp_stress_12}
\end{align}
where the kernel matrix $\mathbf{K}$ is:
\begin{align}
\mathbf{K}(p, x_2, s) =
\begin{bmatrix}
e^{-\beta_1 x_2} - \eta_1 e^{-\beta_2 x_2} \\[0.3em]
-\gamma_1 e^{-\beta_1 x_2} + \eta_1 \gamma_2 e^{-\beta_2 x_2} \\[0.3em]
e^{\alpha x_1} \left[ (\beta_1 \gamma_1 \mu_{12}' - ip \mu_{11}') e^{-\beta_1 x_2} - \eta_1 (\beta_2 \gamma_2 \mu_{12}' - ip \mu_{11}') e^{-\beta_2 x_2} \right] \\[0.3em]
e^{\alpha x_1} \left[ (\beta_1 \gamma_1 \mu_{22}' - ip \mu_{12}') e^{-\beta_1 x_2} - \eta_1 (\beta_2 \gamma_2 \mu_{22}' - ip \mu_{12}') e^{-\beta_2 x_2} \right] \\[0.3em]
\mu_{66}' e^{\alpha x_1}\left[ \left( i p \gamma_1 - \beta_1 \right) ( \, e^{-\beta_1 x_2} -
 \, e^{-\beta_2 x_2} )
\right]
\end{bmatrix}.
\end{align}
Introducing the function:
\begin{align}
     E(p,s) &= (\gamma_1 - \eta_1 \, \gamma_2) \, M_1(p,s), \\
    E^{'}(p,s) &= \frac{\sqrt{p^2 + s^2/v_d^2}}{\xi_1 (\gamma_1 - \eta_1 \, \gamma_2)} \left[(-i s \mu_{12}^{'} + \beta_1 \gamma_1 \mu_{22}^{'}) + \eta_1 \, (i s \mu_{12}^{'} - \beta_2 \gamma_2 \mu_{22}^{'})\right] ,
\end{align}
with
\begin{equation}
\begin{aligned}
\xi_1 &= \frac{1}{(\mu_{11}^{'} + \mu_e {H^0}^2 + \sigma_{11}^0)(N_1 + N_2)(\mu_{12}^{'} + \mu_{66}^{'} + \mu_e {H^0}^2)} \\
&\quad \times \Bigg\{ 
\Big[(\mu_{12}^{'} + \mu_{66}^{'} + \mu_e {H^0}^2)\mu_{12}^{'} - \mu_{22}^{'}(\mu_{11}^{'} + \mu_e {H^0}^2 + \sigma_{11}^0)\Big] \\
&\quad \quad \times \Big[ (\mu_{11}^{'} + \mu_e {H^0}^2 + \sigma_{11}^0) - (\mu_{12}^{'} + \mu_e {H^0}^2 - \sigma_{22}^0) N_1 N_2 \Big] \\
&\quad + \mu_{22}^{'} (\mu_{66}^{'} + \sigma_{22}^0) \Big[ (\mu_{11}^{'} + \mu_e {H^0}^2 + \sigma_{11}^0)(N_1^2 + N_2^2 + N_1 N_2) \\
&\quad \quad + (\mu_{12}^{'} + \mu_e {H^0}^2 - \sigma_{22}^0) N_1^2 N_2^2 \Big] 
\Bigg\},
\end{aligned}
\end{equation}

\begin{align*}
    N_{1,2}^2&=\frac{  (A_1^{'}A_2{'}+B_1^{'} +B_2^{'})\pm \sqrt{(A_1^{'}A_2{'}+B_1^{'} +B_2^{'})^2- 4B_1^{'}B_2^{'}}}{2}\\
A_1^{'}&= \frac{-i(\mu_{12}^{'} + \mu_{66}^{'} + \mu_e {H^0}^2)}{\mu_{66}^{'}+\sigma_{22}^0},\:\: A_2{'}=\frac{-i(\mu_{12}^{'} + \mu_{66}^{'} + \mu_e {H^0}^2)}{\mu_{22}^{'} + \mu_e {H^0}^2+\sigma_{22}^0}, \\ B_1{'}&= \frac{\mu_{11}^{'} + \mu_e {H^0}^2 + \sigma_{11}^0}{\mu_{66}^{'}+\sigma_{22}^0},\:\: B_2{'}=\frac{\mu_{66}^{'}+\sigma_{22}^0}{\mu_{22}^{'} + \mu_e {H^0}^2+\sigma_{22}^0}.
\end{align*}
 Here, the dilatational wave velocity along the \(x_1\)-axis is expressed as: 
\[
v_d = \sqrt{\frac{\mu_{11}}{\rho}} \, v_s ,
\]
where \(v_s = \sqrt{\mu_{66}/\rho}\) denotes the shear wave velocity along the principal axes.
By applying the boundary conditions in Eqs.~\eqref{lap_bc_sigma22} and \eqref{lap_bc_u2}, the problem reduces to the following dual integral equation:
\begin{equation}
    \frac{\xi_1 e^{\alpha x}}{2\pi} \int_{-\infty}^{\infty} \frac{E^{'}(p,s)}{\sqrt{p^2 +{s^2}/{v_d^2}}} E(p,s) e^{-i p x_1} \, dp = -\frac{\sigma_0}{s}, \quad \text{for } x_1 < 0,
\end{equation}
\begin{equation}
    -\frac{1}{2\pi} \int_{-\infty}^{\infty} E(p,s) e^{-i p x_1} \, dp = 0, \quad \text{for } x_1 > 0.
\end{equation}
Let $\bar{\sigma}_+(x_1, s)$ denote the Laplace-transformed normal stress along the positive $x_1$-axis, while $\bar{u_2}_{-}(x_1, s)$ corresponds to the Laplace-transformed vertical displacement along the negative $x_1$-axis.Then,
\be
\bar\sigma_{22}(x_1, 0, s) =
\begin{cases}
\bar{\sigma}_+(x_1, s), &  x_1 \ge 0, \\
\displaystyle \frac{-\sigma_0}{s}, &  x_1 < 0,
\end{cases}, \quad
\bar{u_2}(x_1, 0, s) =
\begin{cases}
0, &  x_1 \ge 0, \\
\bar{u_2}_{-}(x_1, s), &  x_1 < 0.
\end{cases}
\quad
\ee
Furthermore, along the line $x_2 = 0$, the Laplace transforms of the normal stress and vertical displacement take the form:
\begin{align}
\frac{\xi_1}{2\pi} \int_{-\infty}^{\infty} \frac{E^{'}(p,s)}{ \sqrt{p^2 + {s^2}/{v_d^2}}} \, E(p,s) \, e^{-i p x_1} \, dp &= \frac{-\sigma_0}{s} e^{-\alpha x_1} H(-x_1) + \bar\sigma_+(x_1,s) e^{-\alpha x_1}, \label{eq49}\\
-\frac{1}{2\pi} \int_{-\infty}^{\infty} E(p,s) e^{-i p x_1} \, dp &=\bar{u_2}_{-}(x_1, s).
\label{eq50}
\end{align}
Here \( H(x_1) \) is the Heaviside unit step function. Applying Fourier transform to Eqs. \eqref{eq49} and \eqref{eq50}:
\begin{align}
\xi_1 \frac{E^{'}(p,s)}{ \sqrt{p^2 + {s^2}/{v_d^2}}} \, E(p,s)& =\frac{\sigma_0}{s}\left( \frac{\alpha+i p}{\alpha^2+p^2}\right) + \Phi_{+}(p), \label{eq51}\\
-E(p,s)&= {u_2}_{-}(p),
\label{eq52}
\end{align}
where 
\be
\Phi_{+}(p)= \int_0^\infty  \bar\sigma_+(x_1,s) e^{-\alpha x_1} e^{i p x_1} dx_1 , \quad {u_2}_{-}(p)= \int_{-\infty}^0 \bar{u_2}_{-}(x_1, s) e^{i p x_1} dx_1.
\label{eq53}
\ee
Based on the general properties of solutions to elastic wave equations,stress \( \bar{\sigma}_+(x_1, s) \) and the displacement \( \bar{u_2}_{-}(x_1, s) \) are known to remain exponentially bounded as $|x_1| \to \infty$.
The Laplace transform of the normal stress component is defined as:
\begin{equation}
\bar{\sigma}_{22}(x_1, x_2, s) = \int_0^{\infty} \sigma_{22}(x_1, x_2, t) \, e^{-s t} \, dt,
\label{laplacee}
\end{equation}
where \( s \in \mathbb{R}^+ \).
To analyze the behavior of \( \bar{\sigma}_{22}(x_1, x_2, s) \) for large \( x_1 \) or \( x_2 \), consider a fixed point \( (x_1, x_2) \) in the upper half-plane. Let \( \hat{t}(x_1, x_2) \) represent the time at which the first wavefront reaches the point $(x_1, x_2)$. By making the change of variable \( t = t_1 + \hat{t} \), Eq.~\eqref{laplacee} becomes:
\begin{equation}
\bar{\sigma}_{22}(x_1, x_2, s) = e^{-s \hat{t}(x_1, x_2)} \int_0^{\infty} \sigma_{22}(x_1, x_2, t_1 + \hat{t}) \, e^{-s t_1} \, dt_1.
\end{equation}
Accordingly, if the integrand exhibits algebraic dependence on $x_1$ and $x_2$, the asymptotic behavior of the Laplace transform for large values of the spatial coordinates is largely controlled by the arrival time $\hat{t}$ of the first wavefront.
For instance, at any point with $x_1 > 0$ and $x_2 = 0$, the Laplace-domain stress $\bar{\sigma}_{+}(x_1, s)$ is primarily governed by a plane wave propagating away from the crack tip at the dilatational wave speed $v_d$.
In this case, the arrival time is approximately \( \hat{t}(x_1, 0) \approx x_1 / v_d \), and the stress behaves asymptotically as:
\begin{equation}
\bar{\sigma}_+(x_1, s) \sim o\left( e^{-s x_1 (1 - \epsilon)/v_d} \right), \quad \text{as} \quad x_1 \to \infty,
\end{equation}
for some small \( \epsilon > 0 \).
Considering the propagation characteristics of the fields, it is expected that the function \( \bar{\sigma}_+(x_1, s) \) remains bounded asymptotically by an algebraic function of \( x_1 \) multiplied by \( e^{-s x_1/(v_d )} \) as \( x_1 \to +\infty \), for any real positive \( s \). Similarly, the function \( \bar{u_2}_{-}(x_1, s) \) is expected to exhibit algebraic behavior as $x_1 \to - \infty$ 
for any real positive $s$. For large negative values of $x_1$, the motion of the crack faces is mainly influenced by plane waves traveling outward from the surfaces, so that the leading term of $\bar{u}_{2-}(x_1, s)$ becomes independent of $x_1$ as $x_1 \to -\infty$.
Furthermore, following the analytical results established by Noble \cite{noble1959methods}, the following bounds and analyticity conditions hold:
\ If \( |\bar{\sigma}_+(x_1, s)| < S_1 e^{\omega_- x_1} \) as \( x_1 \to +\infty \), then the transformed function \( \Phi_{+}(p) \) is analytic in  \( \text{Im}(p) = \omega > \omega_{-}\) and  if \( |\bar{u_2}_{-}(x_1, s)| < S_2 e^{\omega_+ x_1} \) as \( x_1 \to -\infty \), then the transformed function \( {u_2}_{-}(s) \) is analytic in \( \text{Im}(p) = \omega < \omega_+\).

Based on the above analysis, convergence of the first semi-infinite integral holds for \( \text{Im}(p) > \omega_- = -\dfrac{s}{v_d} \), whereas the second integral is convergent for \( \text{Im}(\omega) < \omega_+ = 0 \). Consequently, the Fourier transform expression given in equation~\eqref{eq53} implies that \( \Phi_{+}(p) \) and \( {u_2}_{-}(s) \) are analytic in the regions \( \text{Im}(p) > -\dfrac{s}{v_d} \) and \( \text{Im}(\omega) < 0 \), respectively.

\subsection{Wiener--Hopf technique}
\label{Wiener--Hopf Technique1}
To facilitate the solution of the problem, the coupled equations governing the system are transformed into a Wiener--Hopf type form. By eliminating the term \( E(p,s) \) from equation~\eqref{eq51}, one obtains the following Wiener--Hopf equation:
\begin{equation}
-\frac{\sigma_0}{s} \left( \frac{\alpha + i p}{\alpha^2 + p^2} \right) - \Phi_{+}(p) = \frac{E'(p,s) \, \xi_1}{\sqrt{p^2 + {s^2}/{v_d^2}}} \, u_{2-}(p).
\label{eq57}
\end{equation}
Define the auxiliary function \( \Sigma(p) \) as:
\begin{equation}
\Sigma(p) = \frac{\Sigma_{-}(p)}{\Sigma_{+}(p)} = \frac{E^{'}(p,s)\, \xi_1}{\sqrt{p^2 + {s^2}/{v_d^2}}}.
\label{eq58}
\end{equation}
Then equation~\eqref{eq57} becomes:
\begin{equation}
-\frac{\sigma_0}{s} \left( \frac{\alpha + i p}{\alpha^2 + p^2} \right) \Sigma_{+}(p) - \Phi_{+}(p)\, \Sigma_{+}(p) = \Sigma_{-}(p)\, u_{2-}(p).
\label{eq59}
\end{equation}
Now introduce the function \( L(p) \) as:
\begin{equation}
L(p) = -\frac{\sigma_0}{s} \left( \frac{\alpha + i p}{\alpha^2 + p^2} \right) \Sigma_{+}(p),
\end{equation}
and decompose it into additive components analytic in their respective half-planes:
\begin{align}
L_{-}(p) = -\frac{\sigma_0}{s} \left( \frac{\alpha + i p}{\alpha^2 + p^2} \right) \Sigma_{+}(0), \quad
L_{+}(p) = -\frac{\sigma_0}{s} \left( \frac{\alpha + i p}{\alpha^2 + p^2} \right) \left[ \Sigma_{+}(p) - \Sigma_{+}(0) \right].
\label{eq62}
\end{align}
Substituting these into equation~\eqref{eq59} yields the transformed Wiener--Hopf equation:
\begin{equation}
L_{+}(p) - \Phi_{+}(p)\, \Sigma_{+}(p) = \Sigma_{-}(p)\, u_{2-}(p) - L_{-}(p) = W(p),
\label{eq60}
\end{equation}
where \( W(p) \) represents a function that can be decomposed into parts analytic in the respective half-planes.

 From the preceding analysis, it is evident that analyticity of the first term holds in the upper half of the complex $p$-plane, specifically for \( \text{Im}(p) > \frac{s}{v_d} \), whereas the second term possesses analyticity in the lower half-plane for \( \text{Im}(p) < 0 \).  Consequently, the domains of analyticity overlap. Once $W(p)$ has been obtained,
 the corresponding expressions for \( \Phi_{+}(p) \) and \(u_{2-}(p) \) can subsequently be evaluated. 
Upon simplification, the following expression is obtained:
\begin{equation}
\begin{aligned}
E(p,s) =& 
\frac{-\sqrt{p^2 + {s^2}/{v_d^2}}}
{\xi_1
\left[ \: p^2(\mu_{11}^{'} + \mu_e {H^0}^2 + \sigma_{11}^0) + i p \alpha(\mu_{11}^{'} + \sigma_{11}^0) + \rho^{'} s^2 \right]
\left[ \alpha \mu_{12}^{'} - i p (\mu_{12}^{'} + \mu_{66}^{'} + \mu_e {H^0}^2) \right]
(\beta_1 + \beta_2)
} \times \\
&  \Bigg\{
\Big[
i p {\mu_{12}^{'}}^2 \alpha + p^2 (\mu_{12}^{'} + \mu_{66}^{'} + \mu_e {H^0}^2) \mu_{12}^{'} - \mu_{22}^{'} p^2 (\mu_{11}^{'} + \mu_e {H^0}^2 + \sigma_{11}^0)
- i p \alpha \mu_{22}^{'} (\mu_{11}^{'} + \sigma_{11}^0)
- \rho^{'} s^2 \mu_{22}^{'}
\Big] \\
&\quad \times \Big[
i p \left( p^2 (\mu_{11}^{'} + \mu_e {H^0}^2 + \sigma_{11}^0) + i p \alpha (\mu_{11}^{'} + \sigma_{11}^0) + \rho^{'} s^2 \right) + \left( \alpha \mu_{12}^{'} - i p (\mu_{12}^{'} + \mu_e {H^0}^2 - \sigma_{22}^0) \right) \beta_1 \beta_2
\Big] \\
&\quad + \mu_{22}^{'} (\mu_{66}^{'} + \sigma_{22}^0) \Big[
i p \left( p^2 (\mu_{11}^{'} + \mu_e {H^0}^2 + \sigma_{11}^0) + i p \alpha (\mu_{11}^{'} + \sigma_{11}^0) + \rho^{'} s^2 \right)
(\beta_1^2 + \beta_2^2 - \beta_1 \beta_2) \\
&\quad - \left( \alpha \mu_{12}^{'} - i p (\mu_{12}^{'} + \mu_e {H^0}^2 + \sigma_{22}^0) \right) \beta_1^2 \beta_2^2
\Big]
\Bigg\}
\end{aligned}
\end{equation}
The zeros of $E(p,s)$ occur solely at $\pm i s / v_R$, where $v_R$ represents the Rayleigh wave velocity.
 Define the function
\begin{equation}
\hat{E}(p) = \frac{E^{'}(p,s)}{p^2 + {s^2}/{v_R^2}},
\end{equation}
such that \( \hat{E}(p) \to 1 \) as \( p \to \infty \). This asymptotic behavior is ensured by an appropriate choice of the parameter \( \xi_1 \).
The function \( \hat{E}(p) \) is analytic in the complex \( p \)-plane and possesses neither poles nor zeros in the finite complex domain. By Cauchy's integral formula, \( \hat{E}(p) \) admits the following factorized representation \cite{freund1998dynamic}:
\begin{equation}
\hat{E}_{\pm}(p) = \exp \left\{ \frac{1}{2\pi i} \int_{\Gamma_{\pm}} \frac{\log \hat{E}(z)}{z - p} \, dz \right\},
\end{equation}
where \( \Gamma_{\pm} \) are contours in the complex \( p \)-plane enclosing the branch points, taken in the upper and lower half-planes respectively.
Therefore, Eq.~\eqref{eq58} can be transformed into the factorized form as:
\begin{equation}
\Sigma(p) = \frac{\Sigma_-(p)}{\Sigma_+(p)} = \frac{\hat{E}(p)\, \xi_1\, \left(p^2 + \frac{s^2}{v_R^2} \right)}{ \sqrt{p^2 + \frac{s^2}{v_d^2}} } 
= \frac{ \hat{E}_+(p)\, \hat{E}_-(p)\, \xi_1\, (p + i s / v_R)(p - i s / v_R)}{ \sqrt{p + i s / v_d} \, \sqrt{p - i s / v_d} }.
\end{equation}

with
\begin{equation}
\Sigma_+(p) = \frac{ \sqrt{p - i s / v_d} }{ (p + i s / v_R)\, \hat{E}_+(p) }, \quad 
\Sigma_-(p) = \frac{ \hat{E}_-(p)\, \xi_1\, (p - i s / v_R)}{ \sqrt{p + i s / v_d} }.
\end{equation}
Each side of Eq.~\eqref{eq60} exhibits analyticity in complementary regions of the complex \( p \)-plane: the right-hand side in the lower half-plane and the left-hand side in the upper half-plane. Since they coincide in their common domain of analyticity, the relation defines a single entire function, denoted by \( W(p) \).
 The nature of this entire function can be deduced from its asymptotic behavior as \( |p| \to \infty \), which is closely related to the behavior of the physical fields near the crack tip at \( x_1 = 0 \).

To determine \( W(p) \), the extended form of Liouville’s theorem is employed. Observe that as \( |p| \to \infty \), the factors \( \Sigma_{+}(p) \sim p^{-1/2} \) and \( \Sigma_{-}(p) \sim p^{1/2} \), implying that both \( L_{+}(p) \) and \( L_{-}(p) \) are bounded in their respective domains of analyticity and vanish as \( |p| \to \infty \).

Moreover, from the physical nature of the problem, it is expected that the Laplace-transformed stress \( \bar{\sigma}_+(x_1, s) \) exhibits a square-root singularity near \( x_1 \to 0^+ \), while the displacement \( \bar{u}_2(x_1, s) \) vanishes as \( x_1 \to 0^- \) to preserve continuity across the crack line. Applying the Abelian theorem  \cite{noble1959methods} for Laplace transforms, the following asymptotic relations hold:
\begin{align}
    &\lim_{x_1 \to 0^+} x_1^{1/2} \bar{\sigma}_+(x_1, s) \sim \lim_{p \to \infty} p^{1/2} \Phi_{+}(p), \\
    &\lim_{x_1 \to 0^-} |x_1|^{-q} \bar{u}_{2-}(x_1, s) \sim \lim_{p \to -\infty} |p|^{1+q} u_{2-}(p),
\end{align}
for some \( q > 0 \). This suggests that \( \Phi_{+}(p) \sim p^{-1/2} \) as \( p \to \infty \), and \( u_{2-}(p) \sim p^{-1-q} \) as \( p \to -\infty \). Therefore, the product terms \( \Phi_{+}(p) \Sigma_{+}(p) \) and \( \Sigma_{-}(p) u_{2-}(p) \) both decay to zero as \( |p| \to \infty \) in their respective half-planes.

Given that all terms in Eq.~\eqref{eq60} vanish at infinity and are analytic in their respective half-planes, it follows from Liouville's extended theorem that the entire function \( W(p) \) must be a constant. Furthermore, as \( |p| \to \infty \), all terms decay to zero, implying that the constant is zero. Hence,$W(p) \equiv 0.$  
Therefore, from Eqs.~\eqref{eq62} and \eqref{eq60}, the expressions for the unknown functions are obtained as:
\begin{align}
\Phi_+(p)&=\frac{L_+(p)}{\Sigma_+(p)}= \frac{\sigma_0}{s} \left( \frac{\alpha + i p}{\alpha^2 + p^2} \right) \left[ \frac{\Sigma_{+}(0)}{\Sigma_{+}(p)}-1\right], \\
u_{2-}(p)&=\frac{L_{-}(p)}{\Sigma_-(p)} = -\frac{\sigma_0}{s} \left( \frac{\alpha + i p}{\alpha^2 + p^2} \right) \frac{\Sigma_{+}(0)}{\Sigma_-(p)}.
\end{align}
\subsection{Stress intensity factor for Mode I loading}
\label{Stress Intensity Factor for Mode I Loading}
To evaluate the stress intensity factor, it is necessary to derive the asymptotic form of the normal stress in the neighborhood of the crack tip. In accordance with Abel’s theorem, the asymptotic correspondence between a function and its Fourier transform is expressed as \cite{noble1959methods}:
\begin{equation}
\lim_{x_1 \to 0^+} \sqrt{x_1} \, f(x_1) = \lim_{p \to \infty} e^{-i \pi/4} \sqrt{\frac{p}{\pi}} \, F(p)
\label{eq69}
\end{equation}
Additionally, the behavior of $\Sigma_+(p)$ as $p \to \infty$ is given by:
\[
\Sigma_+(p) = \frac{1}{p^{1/2}} \quad \text{as} \quad p \to \infty.
\]
Moreover, the continuity condition at the origin requires that:
\[
\sqrt{\hat{E}(0)} = \hat{E}_+(0) = \hat{E}_-(0) =  \frac{v_R}{p} \sqrt{E^{'}(0, s)}.
\]
Based on the preceding analysis, the expression for the \textit{stress intensity factor} in the Laplace domain is given by:
\begin{equation}
\bar K_{\mathrm{I}}(s) = \lim_{x_1 \to 0^+} \sqrt{2\pi x_1} \, \bar{\sigma}_+(x_1, s) = \lim_{p \to +\infty} e^{-i\pi /4} \sqrt{2p} \, \Phi_+(p-i \alpha) = \frac{\sigma_0 \sqrt{2}}{\sqrt{s v_d E^{'}(0,s)}}
\label{SIF1}
\end{equation}
The \textit{dynamic stress intensity factor} in the time domain is obtained by applying the inverse Laplace transform, given by:
\begin{equation}
K_{\mathrm{I}}(t) = \frac{1}{2\pi i} \int_{\text{Br}} \sqrt{2} \, \sigma_0 \,  \frac{e^{s t}}{\sqrt{s} \,\sqrt{v_d E^{'}(0,s)}} \, ds
\label{SIF_time}
\end{equation}
In the next subsection, the Mode~II case is analyzed by following a similar procedure, with modifications corresponding to the shear-mode loading configuration.
\section{Mode-II loading}
\label{Mode-II Loading}
For Mode II (sliding mode) analysis, a sudden application of a uniform shear traction  
$\tau_0$ is considered along the crack faces at time $t = 0$, as schematically illustrated  
in Fig.~\ref{slide2}. Due to the symmetry of the configuration, the analysis can be restricted to the  
upper half-space ($x_2 \ge 0$). The boundary conditions along the crack plane ($x_2 = 0$)  
are specified as follows:
\begin{enumerate}
    \item \textbf{Shear stress condition along the crack plane ($x_1 < 0$):}  
    At the instant $t = 0$, a sudden and uniformly distributed shear traction of magnitude $\tau_0$ is imposed along the crack surfaces. This condition can be mathematically represented as: 
    \begin{equation}
        \sigma_{12}(x_1, 0, t) = -\tau_0 H(t), \quad x_1 < 0,
        \label{eq:modeII_bc1}
    \end{equation}
   where \(H(t)\) denotes the Heaviside unit step function,
 signifying that the load is suddenly applied  
    and remains constant thereafter. The negative sign indicates the chosen sign convention  
    for the shear direction.

    \item \textbf{Normal stress condition along the interface ($x_1$ unrestricted):}  
    No tensile or compressive normal stress acts on the crack plane, hence: 
    \begin{equation}
        \sigma_{22}(x_1, 0, t) = 0, \quad \text{for all } x_1.
        \label{eq:modeII_bc2}
    \end{equation}

    \item \textbf{Tangential displacement continuity ahead of the crack tip ($x_1 > 0$):}  
    For $x_1 > 0$, the interface is assumed to be perfectly bonded. Hence, no relative tangential displacement occurs across the interface, which leads to the condition:  
\begin{equation}
        u_1(x_1, 0, t) = 0, \quad x_1 > 0.
        \label{eq:modeII_bc3}
    \end{equation}
\end{enumerate}

In addition, zero initial conditions are assumed for all field variable. Applying the Laplace transform with respect to time to the boundary conditions \eqref{eq:modeII_bc1}–\eqref{eq:modeII_bc3},  the transformed boundary conditions in the Laplace domain are given by:
\begin{align}
\bar{\sigma}_{12}(x_1, 0, s) &= -\frac{\tau_0}{s}, && \text{for } x_1 < 0, \label{eq:laplace_bc1} \\
\bar{\sigma}_{22}(x_1, 0, s) &= 0, && \text{for all } x_1, \label{eq:laplace_bc2} \\
\bar{u}_1(x_1, 0, s) &= 0, && \text{for } x_1 > 0, \label{eq:laplace_bc3}
\end{align}
Using the boundary condition \eqref{eq:laplace_bc2}, the following relation is obtained:
\begin{equation}
M_2(p,s) = -\eta_2 M_1(p,s), \quad \text{with} \quad \eta_2 = \frac{i p \mu_{12}^{\prime} - \beta_1 \gamma_1 \mu_{22}^{\prime}}{i p \mu_{12}^{\prime} - \beta_2 \gamma_2 \mu_{22}^{\prime}}.
\end{equation}
Thus, the transformed displacement and stress components are obtained as follows:
\begin{align}
\begin{bmatrix}
\bar{u}_1 (x_1, x_2, s) \\[2pt]
\bar{u}_2 (x_1, x_2, s)\\[2pt]
\bar{\sigma}_{11}(x_1, x_2, s) \\[2pt]
\bar{\sigma}_{22} (x_1, x_2, s)\\[2pt]
\bar{\sigma}_{12}(x_1, x_2, s)
\end{bmatrix}
&=
\frac{1}{2\pi} \int_{-\infty}^{\infty} 
\mathbf{\bar{K}}(p,x_2) \,
M_1(p,s) \, e^{-i p x_1} \, dp,
\label{eq:matrix_form}
\end{align}
where the kernel vector $\mathbf{\bar{K}}(p,x_2)$ is given by:
\begin{align}
\mathbf{\bar{K}}(p,x_2) =
\begin{bmatrix}
e^{-\beta_1 x_2} - \eta_2 e^{-\beta_2 x_2} \\[2pt]
- \gamma_1 e^{-\beta_1 x_2} + \eta_2 \gamma_2 e^{-\beta_2 x_2} \\[2pt]
e^{\alpha x_1} \left[ (\beta_1 \gamma_1 \mu_{12}' - i p \mu_{11}') e^{-\beta_1 x_2} 
- \eta_2 (\beta_2 \gamma_2 \mu_{12}' - i p \mu_{11}') e^{-\beta_2 x_2} \right] \\[2pt]
e^{\alpha x_1} \left[ (\beta_1 \gamma_1 \mu_{22}' - i p \mu_{12}') 
\left( e^{-\beta_1 x_2} - e^{-\beta_2 x_2} \right) \right] \\[2pt]
\mu_{66}' e^{\alpha x_1} \left[ (i p \gamma_2 - \beta_2) e^{-\beta_1 x_2} 
- \eta_2 (i p \gamma_1 - \beta_1) e^{-\beta_2 x_2} \right]
\end{bmatrix}.
\label{eq:kernel}
\end{align}
Using the boundary conditions \eqref{eq:laplace_bc1} and \eqref{eq:laplace_bc3}, a system of dual integral equations for the function \( F(p,s) \) is derived:
\begin{align}
\frac{\mu_{66}^{\prime} e^{\alpha x_1}}{2\pi} \, \xi_2 \int_{-\infty}^{\infty} \frac{F^{\prime}(p,s)}{\sqrt{p + s^2/v_s^2}} \, F(p,s) \, e^{-i p x_1} \, dp &= -\frac{\tau_0}{s}, &&  \text{for }x_1 < 0, \label{eq:dual_integral1} \\
\frac{1}{2\pi} \int_{-\infty}^{\infty} F(p,s) \, e^{-i p x_1} \, dp &= 0, && \text{for } x_1 > 0. \label{eq:dual_integral2}
\end{align}
where
\begin{equation}
F(p,s)= (1-\eta_2) M_1(p,s), \quad 
F^{'}(p,s)=\frac{\sqrt{p^2 + s^2/v_s^2}}{(1 - \eta_2) \xi_2} \left( i p \gamma_2 - \beta_2 \right) - \eta_2 \left( i p \gamma_1 - \beta_1 \right),
\end{equation}
with 
\begin{equation}
\begin{aligned}
\xi_2 &= \frac{1}{(N_1 + N_2)\:\mu_{22}^{'}} 
\Bigg[  
\frac{\mu_{12}^{'} + N_1 N_2 \mu_{22}^{'}}{N_1 N_2} 
\left( \frac{\mu_{11}^{'} + \mu_e {H^0}^2 + \sigma_{11}^0}{\mu_{66}^{'} + \sigma_{22}^0} + N_1 N_2 \right) \\
&\quad + \frac{\mu_{12}^{'}(\mu_{12}^{'} + \mu_e {H^0}^2 + \mu_{66}^{'})}{\mu_{66}^{'} + \sigma_{22}^0} \\
&\quad + \Bigg( \left( \frac{\mu_{11}^{'} + \mu_e {H^0}^2 + \sigma_{11}^0}{\mu_{66}^{'} + \sigma_{22}^0} \right)^2 
+ N_1^2 N_2^2 
- \frac{\mu_{11}^{'} + \mu_e {H^0}^2 + \sigma_{11}^0}{\mu_{66}^{'} + \sigma_{22}^0} (N_1^2 + N_2^2) \Bigg) \\
&\quad \times \frac{\mu_{22}^{'} (\mu_{66}^{'} + \sigma_{22}^0)}{(\mu_{12}^{'} + \mu_e {H^0}^2 + \mu_{66}^{'}) N_1 N_2}
\Bigg]
\end{aligned}
\end{equation}
Let $\bar{u_1}_{-}(x_1, s)$ denote the Laplace transform of the horizontal displacement for $x_1 < 0$, and 
$\bar{\tau}_{+}(x_1, s)$ denote the Laplace transform of the shear stress for $x_1 > 0$. 
The boundary conditions along $x_2 = 0$ can then be written as
\begin{align}
\bar{u}_1(x_1,0,s) &=
\begin{cases}
0, & x_1 \ge 0, \\[2pt]
\bar{u_1}_{-}(x_1, s), & x_1 < 0,
\end{cases},
\quad
\bar{\sigma}_{xy}(x_1,0,s) =
\begin{cases}
\bar{\tau}_{+}(x_1,s), & x_1 \ge 0, \\[2pt]
-\dfrac{\tau_0}{s}, & x_1 < 0.
\end{cases}
\label{eq:boundary_conditions}
\end{align}
The Laplace-transformed shear stress and horizontal displacement along \( x_2 = 0 \) are expressed as:
\begin{align}
\frac{\xi_2}{2\pi} \int_{-\infty}^{\infty} \frac{ \mu_{66}^{'}F^{'}(p,s)}{ \sqrt{p^2 + {s^2}/{v_s^2}}} \, F(p,s) \, e^{-i p x_1} \, dp &= \frac{-\tau_0}{s} e^{-\alpha x_1} H(-x_1) + \bar\tau_+(x_1,s) e^{-\alpha x_1}, \label{eq90}\\
\frac{1}{2\pi} \int_{-\infty}^{\infty} F(p,s) e^{-i p x_1} \, dp &=\bar{u_1}_{-}(x_1, s).
\label{eq91}
\end{align}
Applying the Fourier transform to \eqref{eq90}–\eqref{eq91} yields:
\begin{align}
\xi_2 \frac{\mu_{66}^{'} F^{'}(p,s)}{ \sqrt{p^2 + {s^2}/{v_s^2}}} \, F(p,s)& =\frac{\tau_0}{s}\left( \frac{\alpha+i p}{\alpha^2+p^2}\right) + \Xi_{+}(p), \label{eq92}\\
F(p,s)&= {u_1}_{-}(p),
\label{eq93}
\end{align}
where 
\be
\Xi_{+}(p)= \int_0^\infty  \bar\tau_+(x_1,s) e^{-\alpha x_1} e^{i p x_1} dx_1 , \quad {u_1}_{-}(p)= \int_{-\infty}^0 \bar{u_1}_{-}(x_1, s) e^{i p x_1} dx_1.
\label{eq94}
\ee
\subsection{Wiener–Hopf technique}
\label{Wiener--Hopf Technique2}
For the implementation of the Wiener–Hopf technique, the auxiliary function $F(p,s)$ is initially removed from Eq.~\eqref{eq92}.
 This yields a scalar Wiener–Hopf equation of the form:
\begin{align}
\frac{\tau_0}{s}\left( \frac{\alpha+i p}{\alpha^2+p^2}\right) + \Xi_{+}(p) = \xi_2 \frac{\mu_{66}^{'} F^{'}(p,s)}{ \sqrt{p^2 + {s^2}/{v_s^2}}} \,{u_1}_{-}(p) ,
\label{eq95}
\end{align}
where $\Xi_{+}(p)$ and ${u_1}_{-}(p)$ are unknown functions that can be determined using the Wiener–Hopf technique.
Consider a function
\be
\Theta(p)=\frac{\Theta_-(p)}{\Theta_+(p)}=- \xi_2 \frac{\mu_{66}^{'} F^{'}(p,s)}{ \sqrt{p^2 + {s^2}/{v_s^2}}}.
\label{eq96}
\ee
Thus, Eq.~\eqref{eq95} results in the following form:
\be
-\frac{\tau_0}{s}\left( \frac{\alpha+i p}{\alpha^2+p^2}\right) \Theta_+(p) - \Xi_{+}(p) \: \Theta_+(p) =\Theta_-(p) \:{u_1}_{-}(p) .
\ee
Define 
\be
D(p)= D_+(p)+D_-(p)=-\frac{\tau_0}{s}\left( \frac{\alpha+i p}{\alpha^2+p^2}\right) \Theta_+(p),
\ee
 where
 \be
 D_-(p)= -\frac{\tau_0}{s}\left( \frac{\alpha+i p}{\alpha^2+p^2}\right) \Theta_+(0), \quad  D_+(p)= -\frac{\tau_0}{s}\left( \frac{\alpha+i p}{\alpha^2+p^2}\right) [\Theta_+(p)- \Theta_+(0)].
 \ee
 The functions $\Theta_-(p)$ and $\Theta_+(p)$ are given by:
\be
\Theta_-(p)=- \xi_2 \frac{\mu_{66}^{'} (p-i {s}/{v_R}) \hat{F}_-(p)}{ \sqrt{p-i {s}/{v_s}}}, \quad  \Theta_+(p)= \frac{ \sqrt{p+i {s}/{v_s}}}{(p+i {s}/{v_R})\hat{F}_+(p)}, 
\ee
and the function $\hat{F}(p)$ is factorized as:
\begin{equation}
\hat{F}(p) = \hat{F}_+(p) \hat{F}_-(p) = \frac{F'(p, s)}{{p^2 + s^2/v_R^2}}.
\end{equation}
It can be shown that 
\[
\hat{F}(p) \to 1 \quad \text{as} \quad p \to \infty.
\]
Applying the Wiener–Hopf technique, as employed in  normal impact analysis, following expression is obtained:
\be
D_+(p) - \Xi_{+}(p) \, \Theta_+(p) = \Theta_-(p) \, {u_1}_{-}(p) - D_-(p) = 0.
\label{eq101}
\ee
Thus, the solution of Eq.~(\ref{eq101}) is:
\begin{align}
\Xi_+(p)= \frac{D_+(p)}{\Theta_+(p)} =\frac{\tau_0}{s}\left( \frac{\alpha+i p}{\alpha^2+p^2}\right) \left[\frac{\Theta_+(0)}{\Theta_+(p)}-1 \right], \\ {u_1}_{-}(p)= \frac{ D_-(p)}{\Theta_-(p)} =-\frac{\tau_0}{s}\left( \frac{\alpha+i p}{\alpha^2+p^2}\right) \frac{\Theta_+(0)}{\Theta_-(p)}.
\end{align}
Also 
\be
\sqrt{\hat{F}(0)}= \hat{F}_+(0)=\hat{F}_-(0)= \frac{v_R}{s} \sqrt{F'(0, s)},
\ee
\be
\Theta_+(0)=\frac{s}{\sqrt{i sv_s } \sqrt{F'(0, s) }}, \quad \Theta_+(p) =\frac{1}{\sqrt{p}} \quad \text{as} \quad p\to \infty.
\ee

\subsection{Stress intensity factor for Mode II loading}
\label{Stress Intensity Factor for Mode II Loading}
Applying Abel's theorem as stated in Eq.~(\ref{eq69}), which provides the asymptotic relation between a function and its Fourier transform, the expression for the propagation of a semi-infinite crack in an orthotropic medium under mode-II loading is given by:
\begin{equation}
\bar K_{\mathrm{II}}(s) = \lim_{x_1 \to 0^+} \sqrt{2\pi x_1} \, \bar{\tau}_+(x_1, s) = \lim_{p \to +\infty} e^{-i\pi /4} \sqrt{2p} \:\: \Xi_+(p - i \alpha) = \frac{\tau_0 \sqrt{2}}{\sqrt{s v_s F'(0, s)}}.
\label{SIF2}
\end{equation}
The \textit{dynamic stress intensity factor} in the time domain can be obtained by applying the inverse Laplace transform as:
\begin{equation}
K_{\mathrm{II}}(t) = \frac{1}{2\pi i} \int_{\text{Br}} \sqrt{2} \, \tau_0 \,  \frac{e^{s t}}{\sqrt{s} \, \sqrt{v_s F'(0,s)}} \, ds.
\label{SIF2_time}
\end{equation}
The stress intensity factors for Mode I and Mode II in the Laplace domain are derived in Eqs.~\eqref{SIF1} and \eqref{SIF2}, which constitute the principal results of the present analysis.
 In the following section, the Laplace-domain expressions are numerically inverted to the time domain through series expansions employing generalized Chebyshev–Laguerre polynomials.
 The temporal evolution of the resulting stress intensity factors is then investigated for varying material parameters, allowing an assessment of the influence of different factors on the propagation of the semi-infinite crack under the action of sudden traction applied at the crack faces. Detailed numerical simulations are subsequently performed to quantify the dynamic response and the effects of material heterogeneity on crack propagation.

\section{Numerical Simulations and graphical representation}
\label{Numerical Simulations and Graphical Representation}
\subsection{Laplace inversion}
\label{Laplace Inversion}
The dynamic stress intensity factors, \(K_{\mathrm{I}}(t)\) and \(K_{\mathrm{II}}(t)\), are evaluated by numerically inverting their Laplace-domain counterparts, \(\bar{K}_{\mathrm{I}}(s)\) and \(\bar{K}_{\mathrm{II}}(s)\), as formulated in Eqs.~\eqref{SIF1} and~\eqref{SIF2}.
The inversion is carried out using the Chebyshev–Laguerre quadrature technique, which is based on generalized Laguerre polynomials as described by Krylov and Skoblya~\cite{krylov1977handbook}.

Let \(L_k^{(\lambda)}(t)\) denote the generalized Laguerre polynomial of degree \(k\) and type \(\lambda > -1\). The inverse Laplace transform of a function \(\bar{g}(s)\), yielding the original time-domain function \(g(t)\), can be approximated by the expansion:
\begin{equation}
g(t) \approx t^{\lambda} \sum_{k=0}^{\infty} a_k \frac{k!}{\Gamma(k+\lambda+1)} L_k^{(\lambda)}(t),
\label{eq:LaguerreInversion}
\end{equation}
where the expansion coefficients \(a_k\) are defined by:
\begin{equation}
a_k = \frac{(-1)^k}{k!} \left. \frac{d^k}{dz^k} \left\{ \frac{1}{z^{\lambda+1}} F\left(\frac{1}{z}\right) \right\} \right|_{z=1}.
\label{eq:ak_coeff}
\end{equation}
The generalized Laguerre polynomial \(L_k^{(\lambda)}(t)\) is given explicitly as:
\begin{equation}
L_k^{(\lambda)}(t) = \sum_{m=0}^{k} \frac{\Gamma(k+\lambda+1)}{\Gamma(m+\lambda+1)} \cdot \frac{(-t)^m}{m! (k-m)!}.
\label{eq:LaguerrePoly}
\end{equation}

This quadrature approach ensures accurate numerical inversion, particularly suited for functions with singular behavior or rapid variations near \(t = 0\), which are typical in fracture mechanics problems involving transient dynamic loading.

\subsection{Graphical analysis of normalized dynamic stress intensity factor}
\label{graphical}
This subsection presents a comprehensive graphical investigation of the normalized dynamic stress intensity factors, \( K_{\mathrm{I}}/\sigma_0 \) and \( K_{\mathrm{II}}/\tau_0 \), evaluated at the crack tip (\( x_1 = 0 \)), by employing a numerical inversion scheme based on standard Laguerre polynomials implemented in \textsc{MATLAB}. The evolution of these intensity factors with time is studied in relation to various physical parameters that influence the crack propagation behavior in magnetoelastic orthotropic media. The parameters under consideration include the heterogeneity parameter \( \alpha \), the magneto-elastic coupling parameter \( \mu_e {H^0}^2/\mu_{66}' \), the initial stress components \( \sigma_{11}^0 \) and \( \sigma_{22}^0 \) applied along the \( x_1 \)- and \( x_2 \)-axes, respectively, and the rotational parameter \( \Omega \), which characterizes the angular velocity of the rotating medium.

The material constants used for numerical simulation are listed in Table~\ref{table:material_constants}. Furthermore, to elucidate the effects of spatially varying properties and anisotropy, the fracture response of functionally graded materials (FGMs) is examined and compared against that of isotropic counterparts for each parametric variation. 
\begin{table}[htbp]
\centering
\caption{Values of elastic constants (in $\times 10^4$ MPa) and densities (in g/cm\textsuperscript{3}) for different materials \cite{alam2025mode,rubio2000dynamic}.}
\label{table:material_constants}
\renewcommand{\arraystretch}{1.3}
\begin{tabular}{|c|c|c|c|c|c|c|}
\hline
\textbf{Name} & $\mathcal{\mu}_{11}^{'}$ & $\mathcal{\mu}_{12}^{'}$ & $\mathcal{\mu}_{22}^{'}$ & $\mathcal{\mu}_{66}^{'}$ & $\rho^{'}$ & \textbf{Type} \\
\hline
$\alpha$- Uranium & 21.47 & 7.43 & 19.86 & 4.65 & 19.07 & Orthotropic \\
\hline
Graphite Epoxy & 15.536 & 0.367 & 1.631 & 0.748 & 1.6& Orthotropic \\
\hline
Steel & $21.98$ & $6.593$ & $21.98$ & $7.692$ & $7.840$ & Isotropic \\
\hline
\end{tabular}
\end{table}

\subsubsection{Time-dependent behavior of the stress intensity factor}
The temporal evolution of the stress intensity factor (SIF), as illustrated in Figs.~\ref{Figure 1}--\ref{Figure 10}, reveals an initial rapid increase followed by a gradual stabilization phase. This response is typical of transient dynamic loading conditions and has been similarly reported in previous investigations, such as the work by Rubio and Ortiz~\cite{rubio2000dynamic}. However, consistent with their conclusions, it must be emphasized that this apparent stabilization does not correspond to convergence toward a true quasi-static or equilibrium value.
The initial surge in the SIF is attributed to the instantaneous application of stress or displacement boundary conditions at \( t = 0 \), which generates elastic stress waves that rapidly propagate through the domain and concentrate at the crack tip. These waves induce a transient amplification of the SIF due to wave-crack tip interactions and local energy accumulation. As time advances, the stress waves experience multiple reflections and dispersions, gradually redistributing energy throughout the medium and reducing the intensity of localized dynamic effects. 
Material-dependent behavior is also evident in the temporal response of the SIF. Among the three materials studied, the isotropic material demonstrates the largest variation in SIF over time, indicating a higher sensitivity to transient loading. In contrast, graphite–epoxy composites display minimal sensitivity under Mode I loading, while $\alpha$-Uranium exhibits the least variation under Mode II conditions. These differences can be attributed to the anisotropic wave propagation characteristics and directional stiffness variations inherent to each material system.
Figs.~\ref{Figure 1}–\ref{Figure 5} correspond to Mode I fracture (opening mode), while Figs.~\ref{Figure 6}–\ref{Figure 10} represent Mode II fracture (in-plane shear). Despite the different loading configurations and resulting displacement fields, both fracture modes exhibit analogous temporal SIF responses. 

\subsubsection{Influence of functional grading on the dynamic stress intensity factor}
Figures~\ref{Figure 1} and \ref{Figure 6} illustrate the influence of material heterogeneity, characterized by the functional grading parameter \( \alpha \), on the temporal evolution of the stress intensity factors (SIFs) for Mode I and Mode II, denoted by \( K_{\mathrm{I}}(t)/\sigma_0 \) and \( K_{\mathrm{II}}(t)/\tau_0 \), respectively.
As observed, an increase in the amplitude of heterogeneity—i.e., larger values of \( \alpha \)—leads to a noticeable enhancement in the magnitudes of both \( K_{\mathrm{I}}(t)/\sigma_0 \) and \( K_{\mathrm{II}}(t)/\tau_0 \). This trend is a direct consequence of the directional material grading imposed along the \( x_1 \)-axis, which modifies the local elastic modulus and wave propagation characteristics around the crack tip.
In the case of {Mode I} loading, which involves tensile opening perpendicular to the crack plane, an increase in \( \alpha \) results in a steeper stiffness gradient. This gradient leads to the refraction and focusing of tensile stress waves toward the crack tip, thereby intensifying the local stress concentration. 
For {Mode II} loading, where in-plane shear is dominant, the spatial variation in shear modulus caused by functional grading induces stronger refraction and reflection of shear waves near the crack tip. This effect results in an increasingly asymmetric and concentrated stress field under higher \( \alpha \), which in turn elevates the shear tractions at the crack tip. Consequently, the dynamic Mode II SIF increases with greater heterogeneity.

\subsubsection{Influence of magneto-elastic coupling parameter on the dynamic stress Intensity factor}
Figures~\ref{Figure 2} and \ref{Figure 7} illustrate the influence of the magneto-elastic coupling parameter($\mu_e {H^0}^2/\mu_{66}'$) on the temporal evolution of the normalized stress intensity factors (SIFs) for Mode I and Mode II, represented by \( K_{\mathrm{I}}(t)/\sigma_0 \) and \( K_{\mathrm{II}}(t)/\tau_0 \), respectively.
The results demonstrate that strengthening the magneto-elastic coupling parameter diminishes the Mode I stress intensity factor, while simultaneously amplifying the Mode II stress intensity factor. This contrasting behavior can be attributed to the directionality of the applied magnetic field and its interaction with the deformation modes.
Specifically, the magnetic field is assumed to act along the \( x_3 \)-direction, whereas crack propagation and deformation occur in the \( x_1x_2 \)-plane.
In this configuration, the Lorentz forces and magneto-mechanical interactions have limited coupling with the normal opening displacements associated with Mode I loading. As a result, the magnetic field introduces stress redistribution and damping effects that reduce tensile stress concentration at the crack tip, thereby lowering \( K_{\mathrm{I}} \), as also observed in prior studies~\cite{alam2025mode}.
In contrast, under Mode II loading, which involves in-plane shear deformation, the applied magnetic field couples more effectively with the shear stress field. The resulting magneto-mechanical interactions amplify asymmetric stress distributions near the crack tip, particularly due to induced Lorentz forces and magnetic anisotropy. This leads to an increase in \( K_{\mathrm{II}} \) with higher magneto-elastic coupling.
Additionally, the magneto-elastic coupling induces anisotropic stiffness modulation, wherein the material exhibits increased resistance to tensile opening but reduced resistance to shear deformation. This further explains the observed suppression of Mode I SIF and the enhancement of Mode II SIF.

\subsubsection{Influence of the rotation parameter on the dynamic stress intensity factor}
Figures~\ref{Figure 3} and \ref{Figure 8} illustrate the effect of the rotation parameter (\( \Omega \)) on the time-dependent behavior of the normalized stress intensity factors (SIFs) for Mode I and Mode II, given by \( K_{\mathrm{I}}(t)/\sigma_0 \) and \( K_{\mathrm{II}}(t)/\tau_0 \), respectively. Increasing the rotation parameter results in a decrease in both Mode I and Mode II SIFs.
 The observed behavior arises from rotational inertia effects, with centrifugal and Coriolis forces modifying the stress and displacement fields around the crack tip in the rotating frame.
 As the angular velocity increases, centrifugal forces introduce radial body forces that act perpendicular to the axis of rotation. For a semi-infinite crack extending along the \( x_1 \)-axis (from \( -\infty \) to 0), these centrifugal effects result in a uniform redistribution of normal stresses along the crack front. This tends to reduce the net tensile stress concentration near the crack tip, thereby lowering the Mode I SIF.
Simultaneously, Coriolis forces—which arise due to particle motion in the rotating frame—act perpendicular to both the rotation axis and the velocity vector. In the case of Mode II fracture, which involves in-plane shear deformation along the \( x_1 \)-axis, the Coriolis force introduces asymmetric in-plane inertial forces. However, for a semi-infinite crack, the lack of a rear crack tip or finite-length boundaries means that wave reflections are absent, and rotational effects accumulate over the infinite crack front. Instead of amplifying shear stress near the tip, these rotational inertial forces interfere with coherent shear wave propagation and weaken the stress gradient, leading to a decrease in the Mode II SIF.
Thus, for a semi-infinite crack configuration, rotation contributes a dynamic stress shielding effect in both tensile and shear modes. The infinite extent of the crack geometry amplifies the uniform redistributive action of centrifugal loading while eliminating stress reflection effects from the far end. As a result, crack-tip energy localization is suppressed, which may delay crack initiation or slow crack growth in rotating structural components aligned with the rotation axis.

\subsubsection{Influence of horizontal initial stress on the dynamic stress intensity factor}
Figures~\ref{Figure 4} and \ref{Figure 9} illustrate the effect of horizontal initial normal stress \( \sigma_{11}^0 \) on the temporal evolution of the normalized stress intensity factors (SIFs) for Mode I and Mode II, represented by \( K_{\mathrm{I}}(t)/\sigma_0 \) and \( K_{\mathrm{II}}(t)/\tau_0 \), respectively.
The results indicate that a rise in the horizontal initial stress \( \sigma_{11}^0 \), especially in compression, enhances the Mode I stress intensity factor.
 The observed trend arises because Mode I fracture is dominated by opening displacements normal to the crack plane.
 The imposed compressive stress in the transverse direction induces additional constraint around the crack tip, thereby intensifying stress localization and increasing the energy available for crack opening. As a result, the dynamic Mode I SIF becomes larger under stronger lateral confinement.
On the other hand, for Mode II fracture, which is governed by in-plane shear displacement, the presence of compressive horizontal stress alters the surrounding stress field in a manner that suppresses relative tangential movement along the crack faces. This geometric confinement effectively reduces the shear stress intensity at the crack tip. Consequently, an increase in \( \sigma_{11}^0 \) leads to a decrease in the Mode II SIF.
This contrasting response between Mode I and Mode II highlights the anisotropic influence of prestress conditions on dynamic fracture behavior, especially in heterogeneous or preloaded solids. The findings underscore the importance of accounting for initial stress fields when assessing crack dynamics in engineering structures subjected to multi-axial loading.

\subsubsection{Influence of vertical initial stress on the dynamic stress intensity factor}

Figures~\ref{Figure 5} and \ref{Figure 10} illustrate the effect of vertical initial normal stress \( \sigma_{22}^0 \) on the temporal evolution of the normalized stress intensity factors (SIFs) for Mode I and Mode II, represented by \( K_{\mathrm{I}}(t)/\sigma_0 \) and \( K_{\mathrm{II}}(t)/\tau_0 \), respectively.
It is observed that increasing the magnitude of the initial vertical compressive stress \( \sigma_{22}^0 \) leads to a reduction in the Mode I SIF and a corresponding increase in the Mode II SIF. 
Under Mode I loading, which involves normal opening of the crack faces, vertical compressive stress directly opposes the opening displacement and acts uniformly along the infinite crack front. This leads to a global suppression of tensile energy concentration at the tip and causes a marked reduction in \( K_{\mathrm{I}}(t)/\sigma_0 \). In contrast, Mode II loading induces in-plane shear displacement along the crack surfaces. The presence of vertical compressive stress enhances the lateral constraint perpendicular to the crack path. For a semi-infinite crack, this promotes greater shear distortion along the faces, intensifying the local stress gradients and resulting in an elevated \( K_{\mathrm{II}} \). 
These results reflect the fundamental behavior of dynamic fracture in semi-infinite configurations, where directional pre-stress fields exert asymmetric influence on the fracture modes due to the absence of tip reflections and the continuous stress wave accumulation near the advancing front.

\bfg[htbp]
\centering
\begin{subfigure}[b] {0.8\textwidth}
\includegraphics[width=\textwidth ]{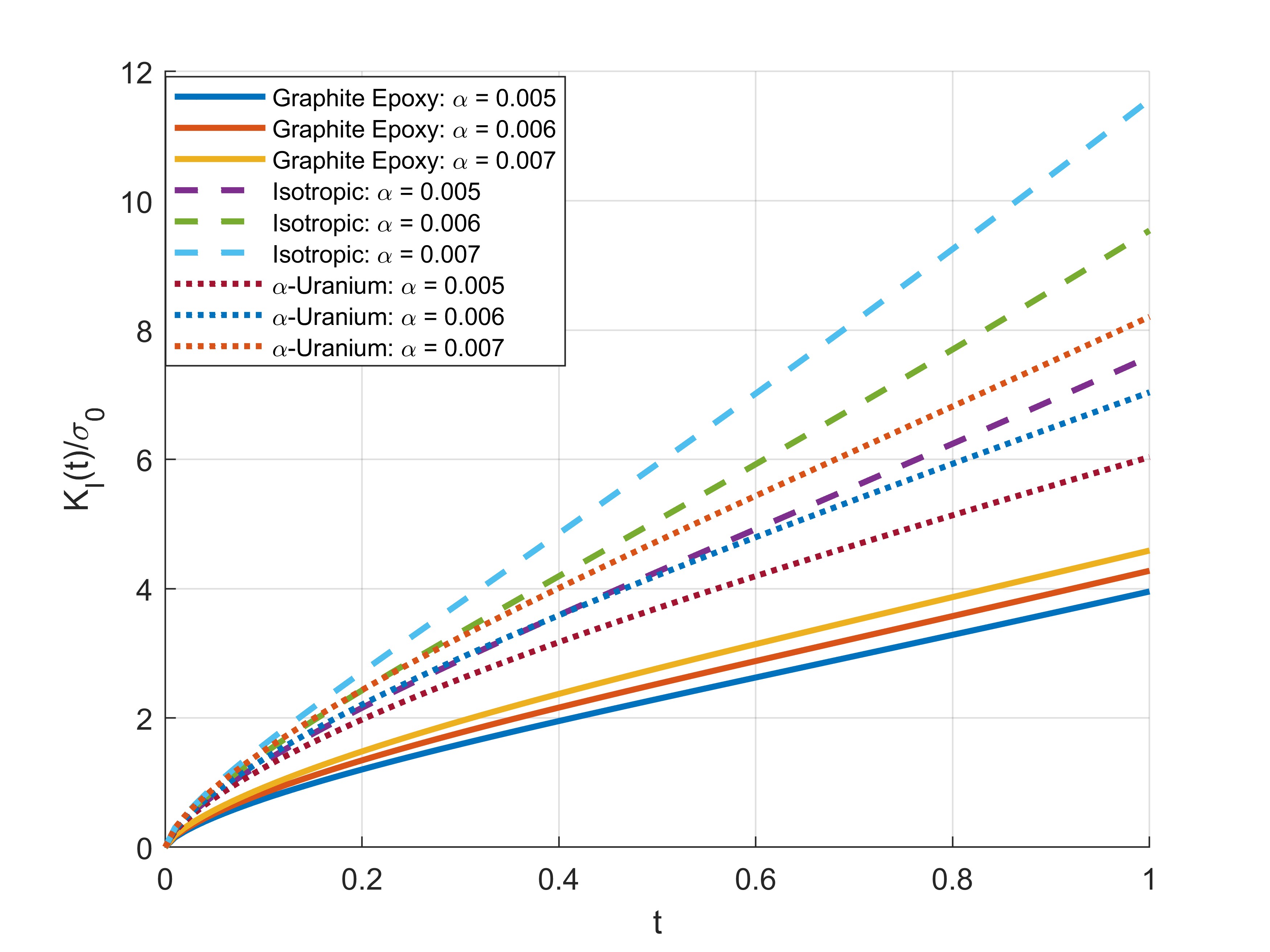}
\caption{}
\label{Fig_1}
\end{subfigure}
~
\begin{subfigure}[b] {0.8\textwidth}
\includegraphics[width=\textwidth ]{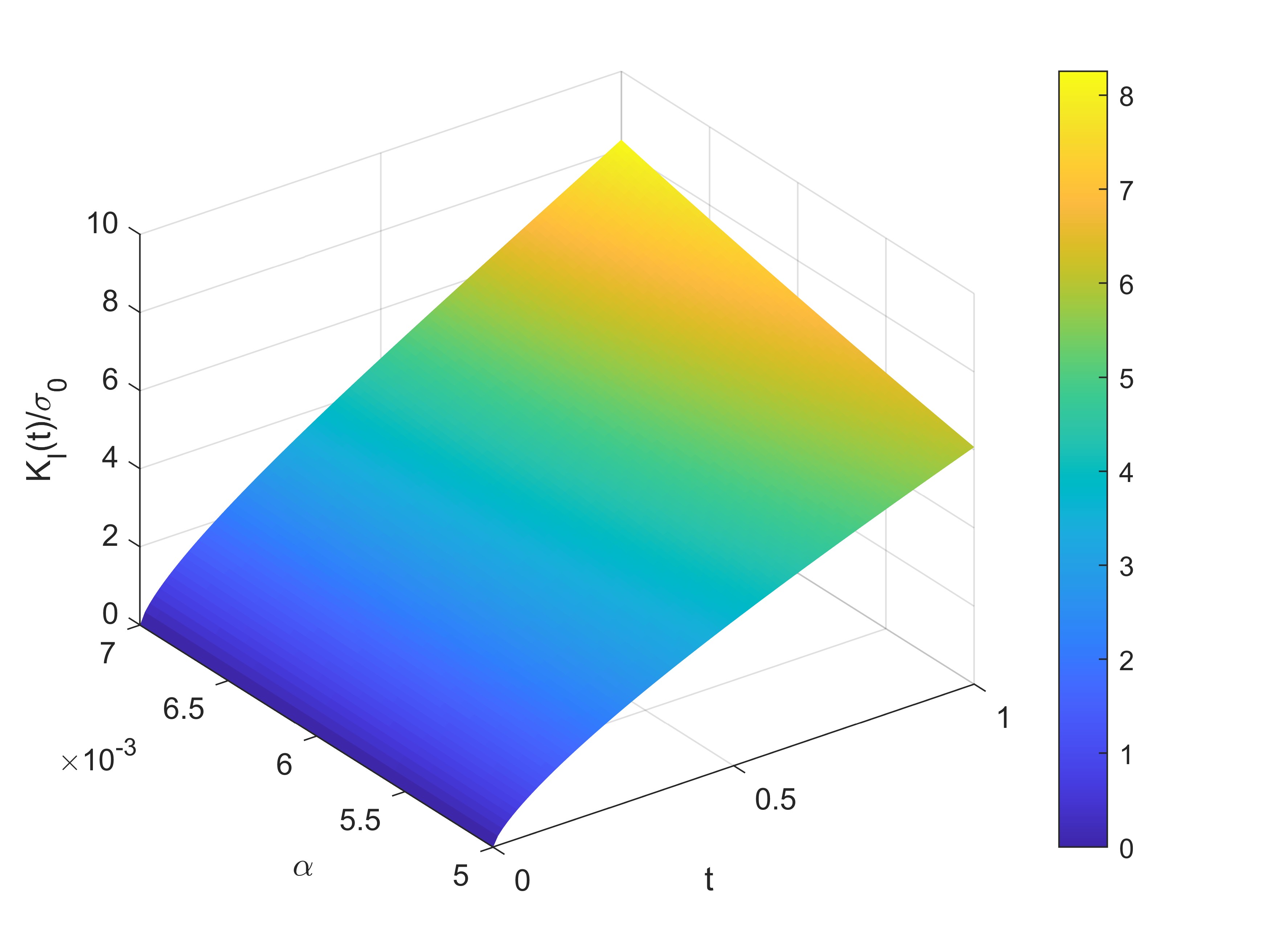}
\caption{}
\label{Fig_2}
\end{subfigure}

\caption{Normalized stress intensity factor ($K_{\mathrm{I}}/\sigma_0$) at $x_1 = 0$ of a semi-infinite crack versus time, illustrating the influence of heterogeneity parameter ($\alpha$).}
\label{Figure 1}
\efg

\bfg[htbp]
\centering
\begin{subfigure}[b] {0.8\textwidth}
\includegraphics[width=\textwidth ]{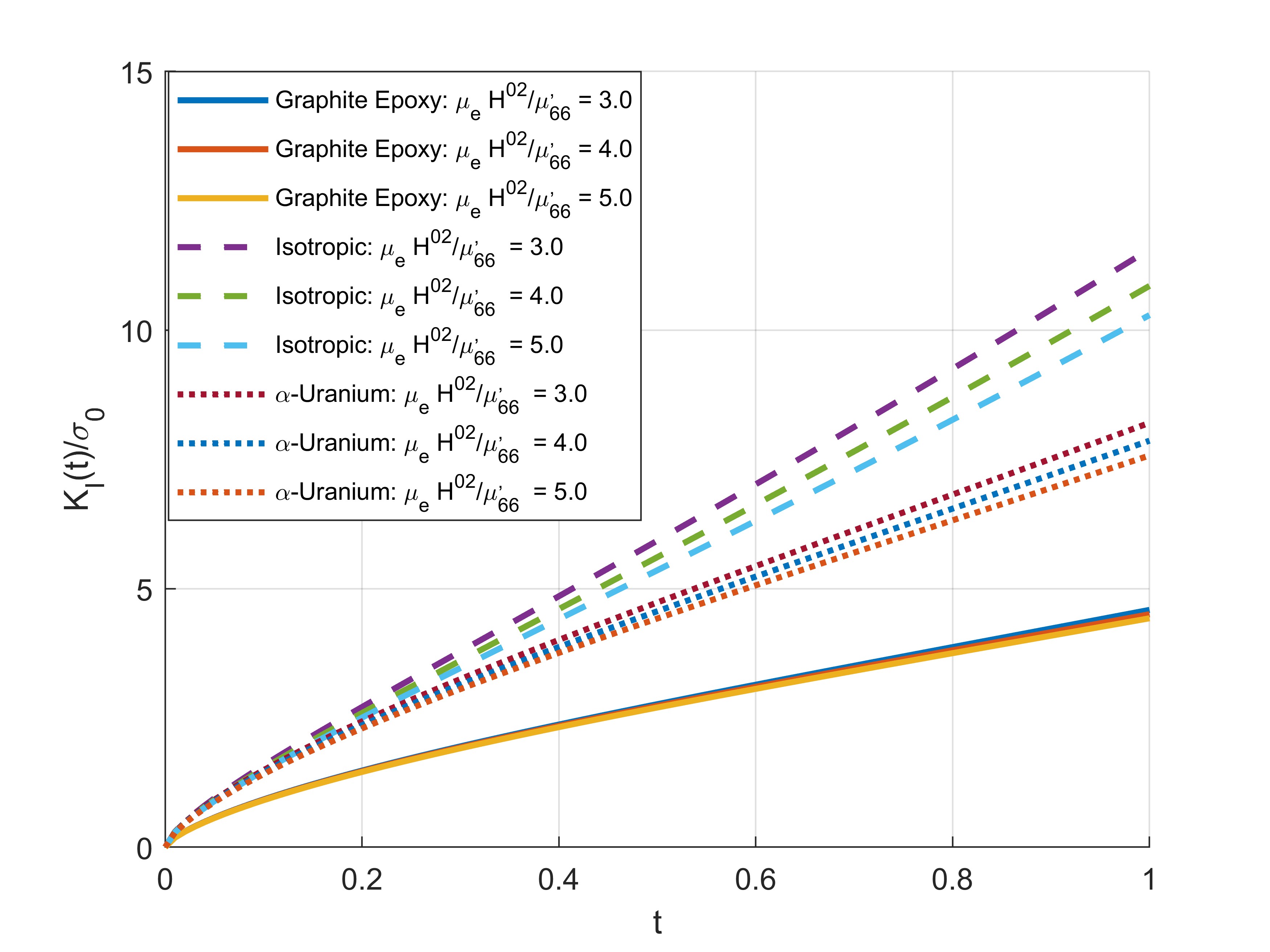}
\caption{}
\label{Fig_3}
\end{subfigure}
~
\begin{subfigure}[b] {0.8\textwidth}
\includegraphics[width=\textwidth ]{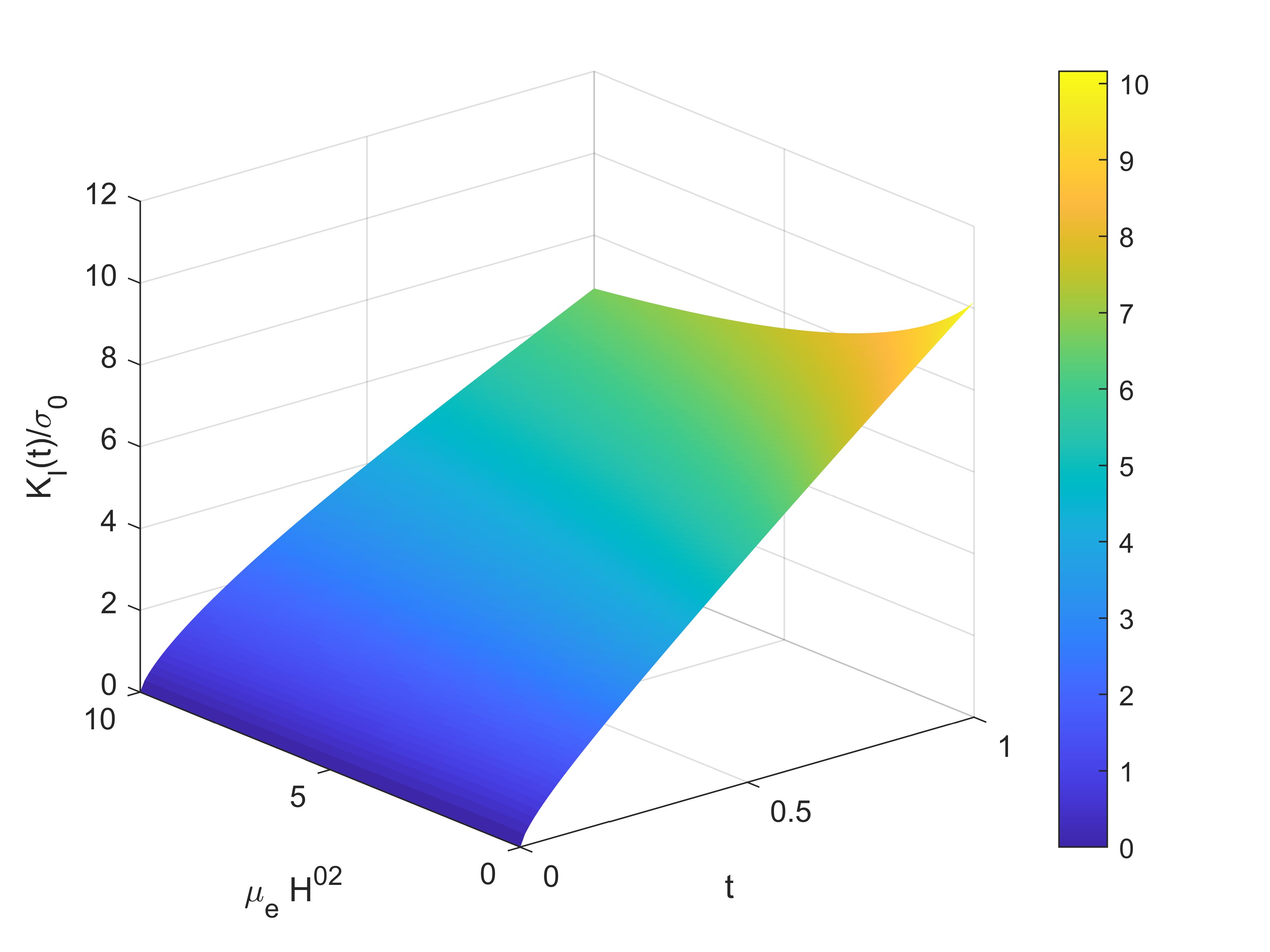}
\caption{}
\label{Fig_4}
\end{subfigure}

\caption{Normalized stress intensity factor ($K_{\mathrm{I}}/\sigma_0$) at $x_1 = 0$ of a semi-infinite crack versus time, illustrating the influence of magneto-elastic coupling parameter ($\mu_e {H^0}^2/\mu_{66}'$).}
\label{Figure 2}
\efg

\bfg[htbp]
\centering
\begin{subfigure}[b] {0.8\textwidth}
\includegraphics[width=\textwidth ]{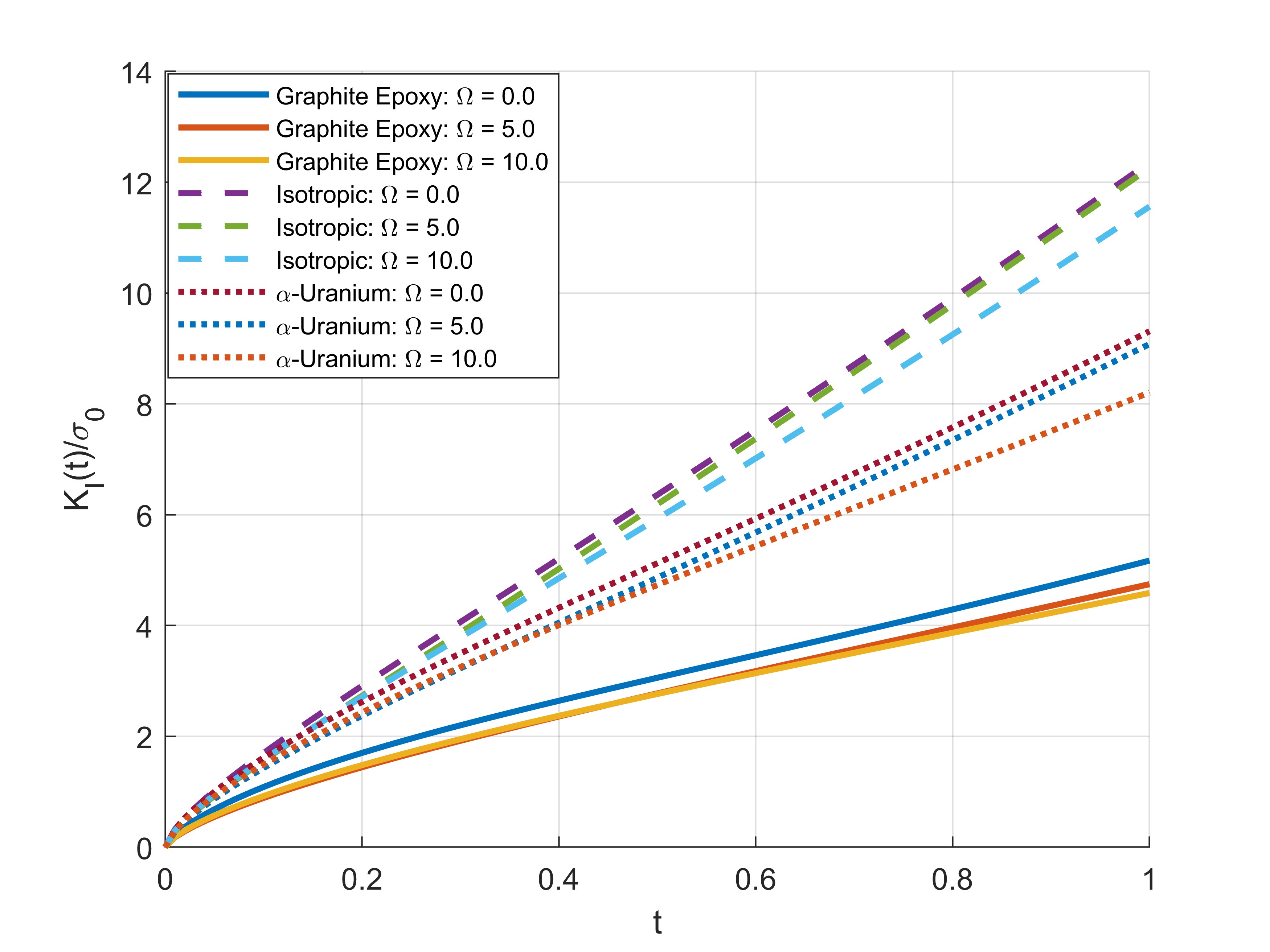}
\caption{}
\label{Fig_5}
\end{subfigure}
~
\begin{subfigure}[b] {0.8\textwidth}
\includegraphics[width=\textwidth ]{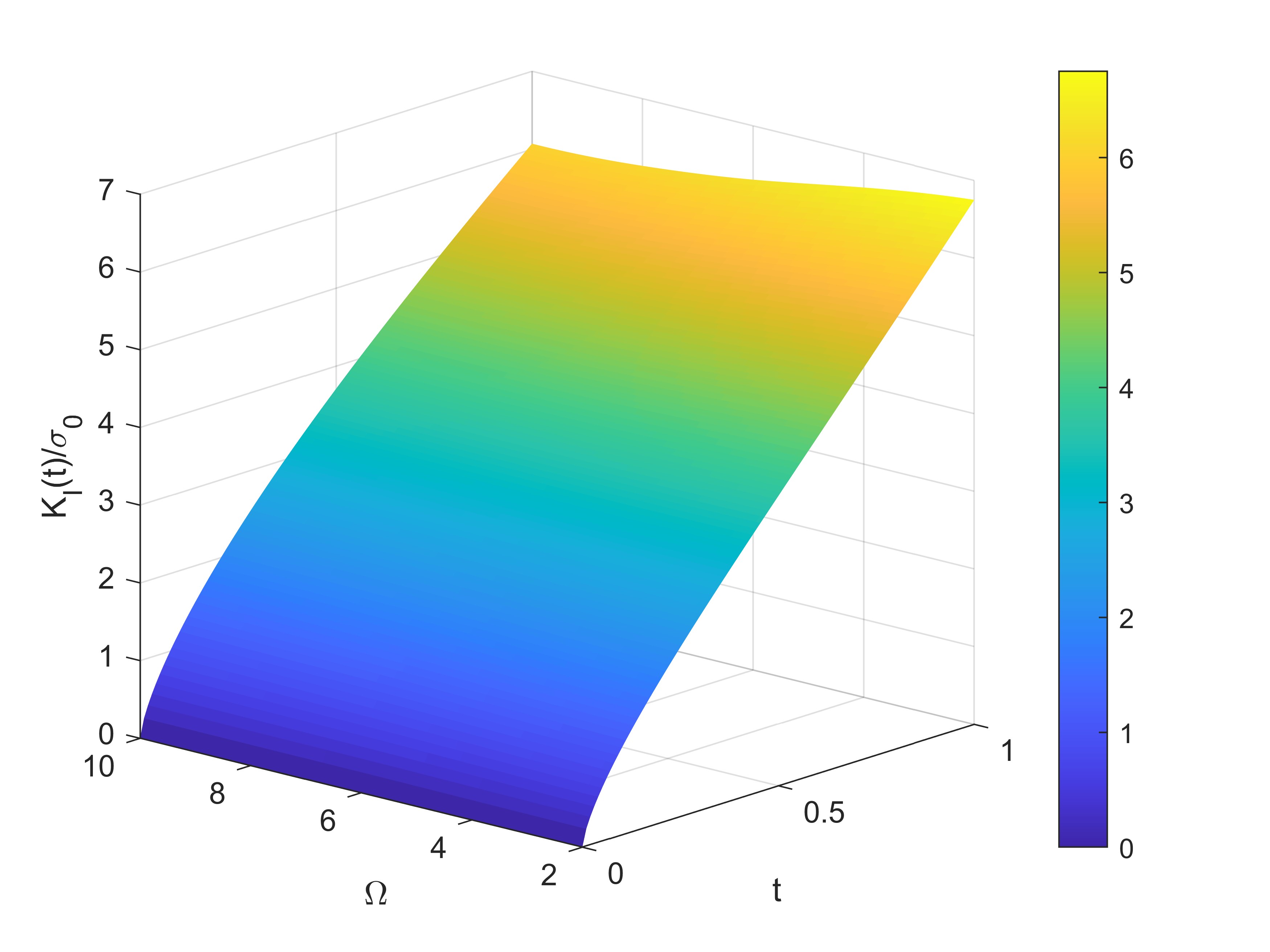}
\caption{}
\label{Fig_6}
\end{subfigure}

\caption{Normalized stress intensity factor ($K_{\mathrm{I}}/\sigma_0$) at $x_1 = 0$ of a semi-infinite crack versus time, illustrating the influence of rotational parameter ($\Omega$).}
\label{Figure 3}
\efg

\bfg[htbp]
\centering
\begin{subfigure}[b] {0.8\textwidth}
\includegraphics[width=\textwidth ]{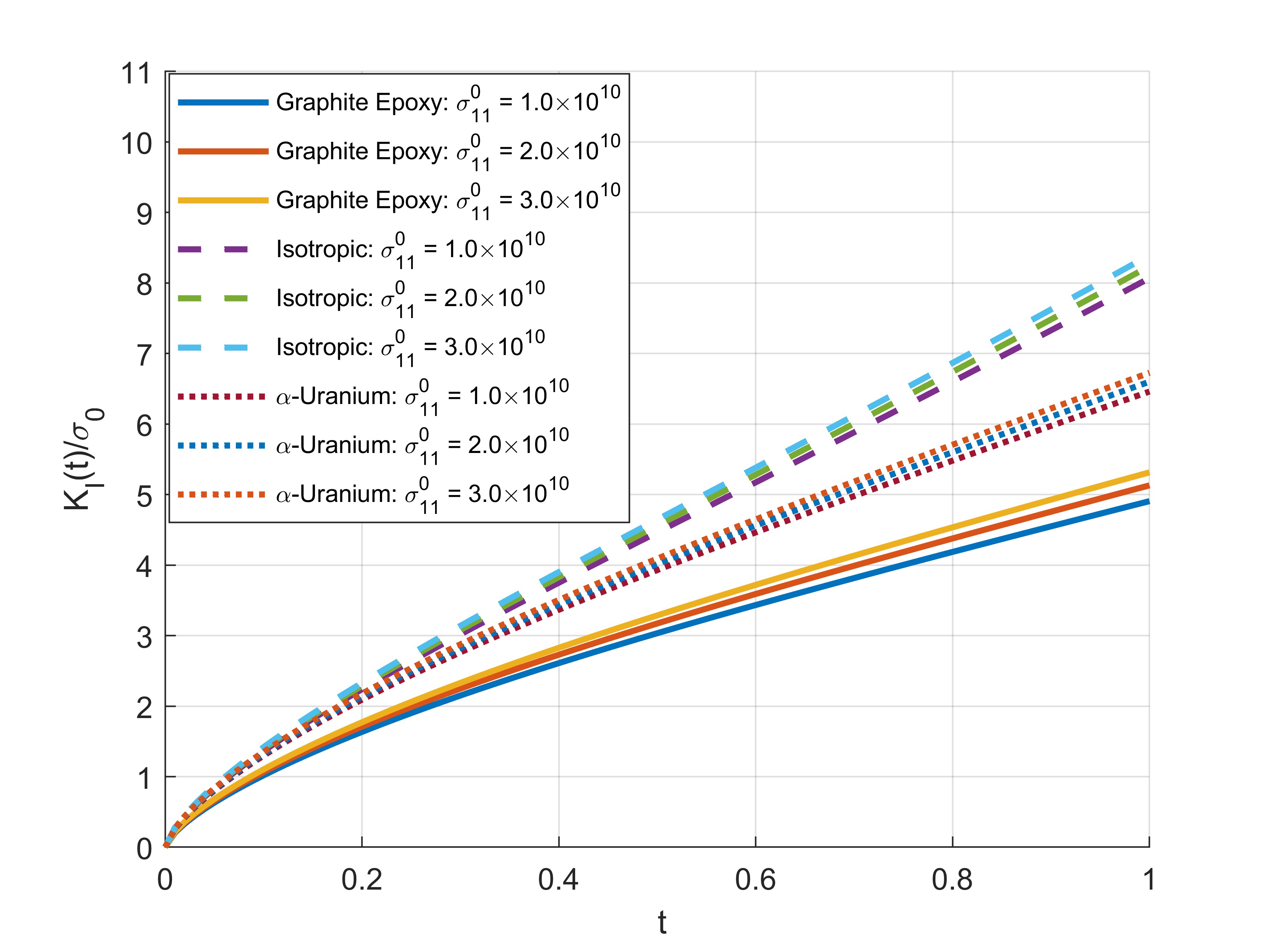}
\caption{}
\label{Fig_7}
\end{subfigure}
~
\begin{subfigure}[b] {0.8\textwidth}
\includegraphics[width=\textwidth ]{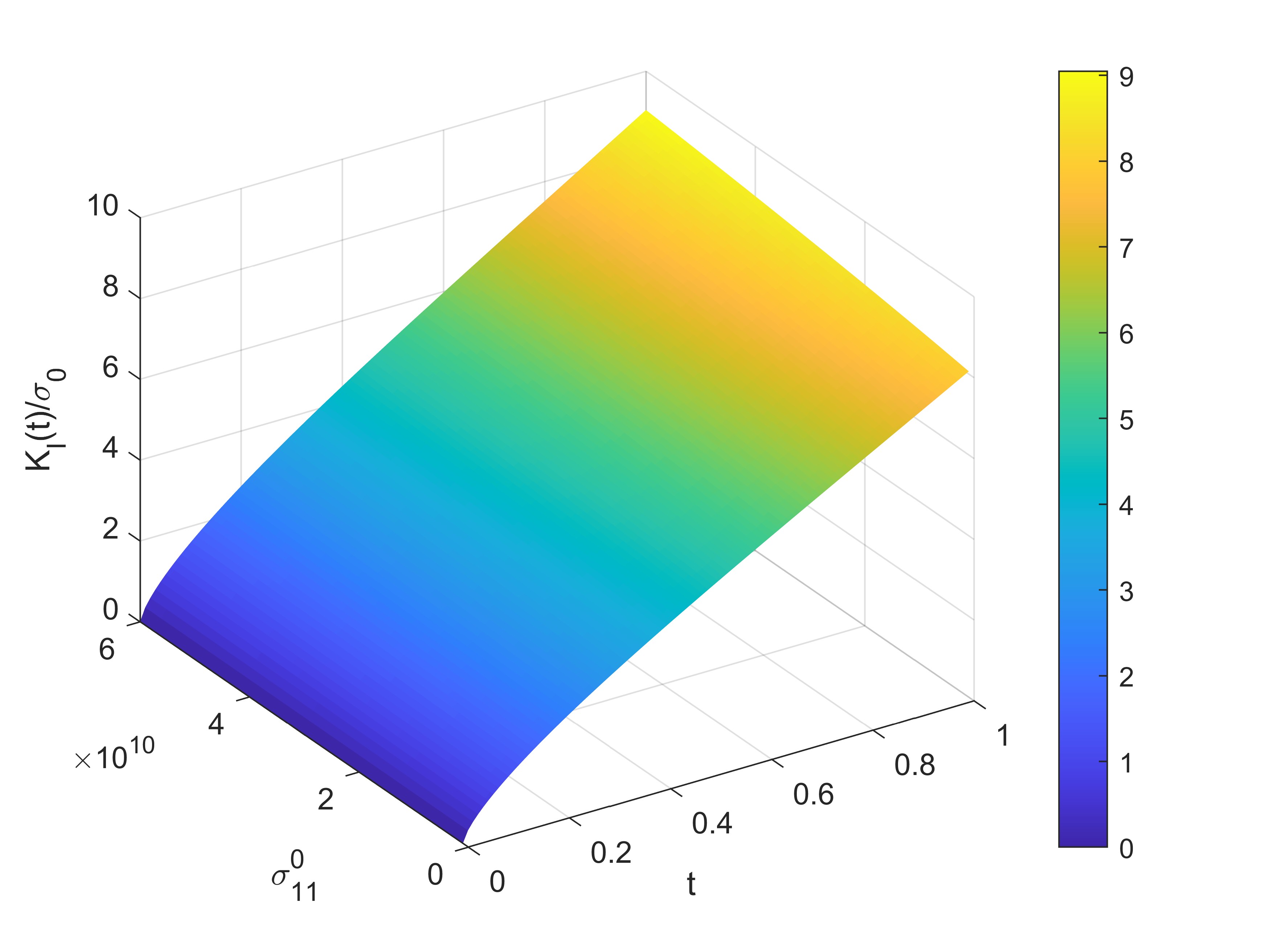}
\caption{}
\label{Fig_8}
\end{subfigure}

\caption{Normalized stress intensity factor ($K_{\mathrm{I}}/\sigma_0$) at $x_1 = 0$ of a semi-infinite crack versus time, illustrating the influence of initial stress along the $x_1$-axis ($\sigma_{11}^0$).}
\label{Figure 4}
\efg
\bfg[htbp]
\centering
\begin{subfigure}[b] {0.8\textwidth}
\includegraphics[width=\textwidth ]{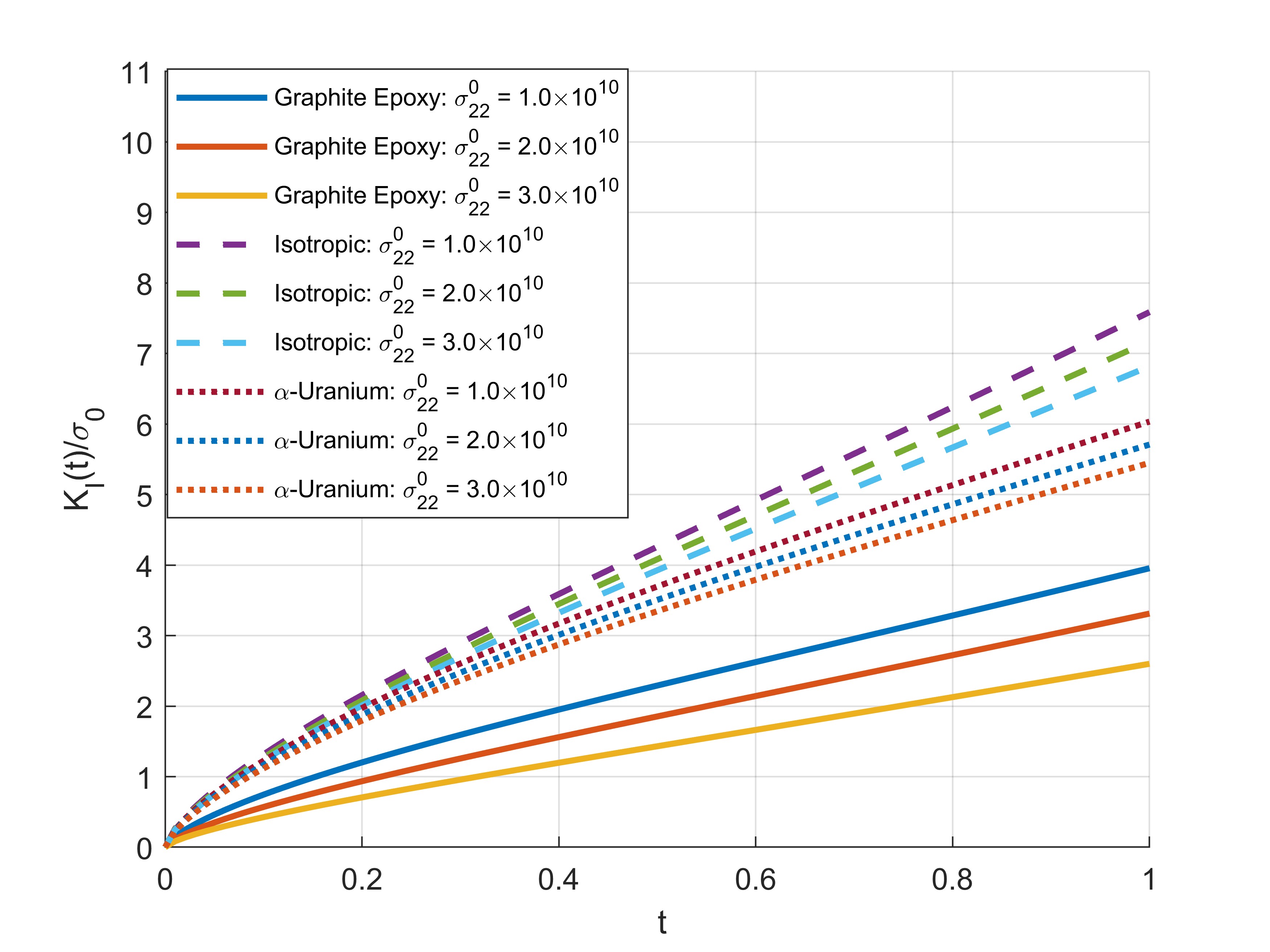}
\caption{}
\label{Fig_9}
\end{subfigure}
~
\begin{subfigure}[b] {0.8\textwidth}
\includegraphics[width=\textwidth ]{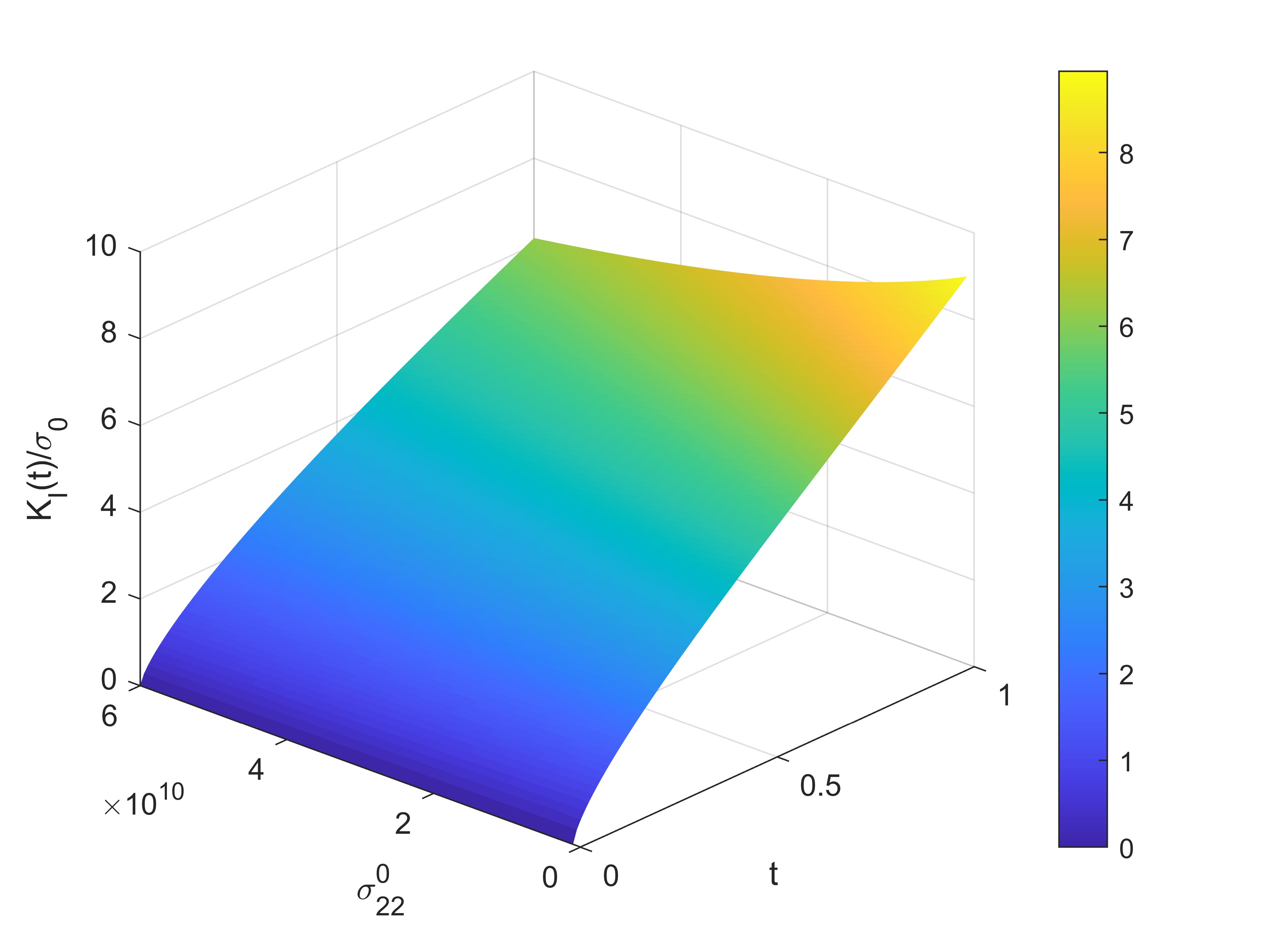}
\caption{}
\label{Fig_10}
\end{subfigure}

\caption{Normalized stress intensity factor ($K_{\mathrm{I}}/\sigma_0$) at $x_1 = 0$ of a semi-infinite crack versus time, illustrating the influence of initial stress along the $x_2$-axis ($\sigma_{22}^0$).}
\label{Figure 5}
\efg
\bfg[htbp]
\centering
\begin{subfigure}[b] {0.8\textwidth}
\includegraphics[width=\textwidth ]{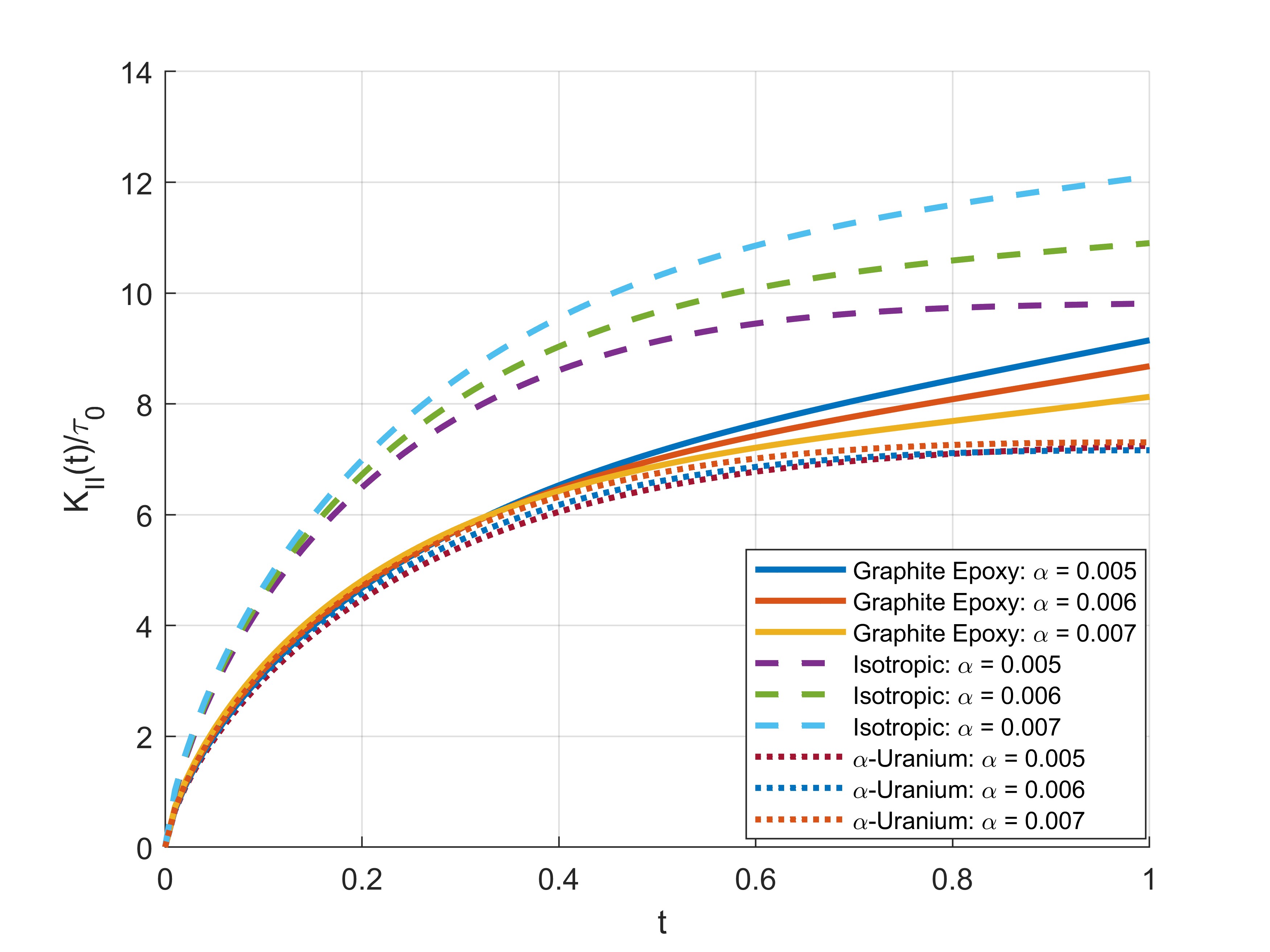}
\caption{}
\label{Fig_11}
\end{subfigure}
~
\begin{subfigure}[b] {0.8\textwidth}
\includegraphics[width=\textwidth ]{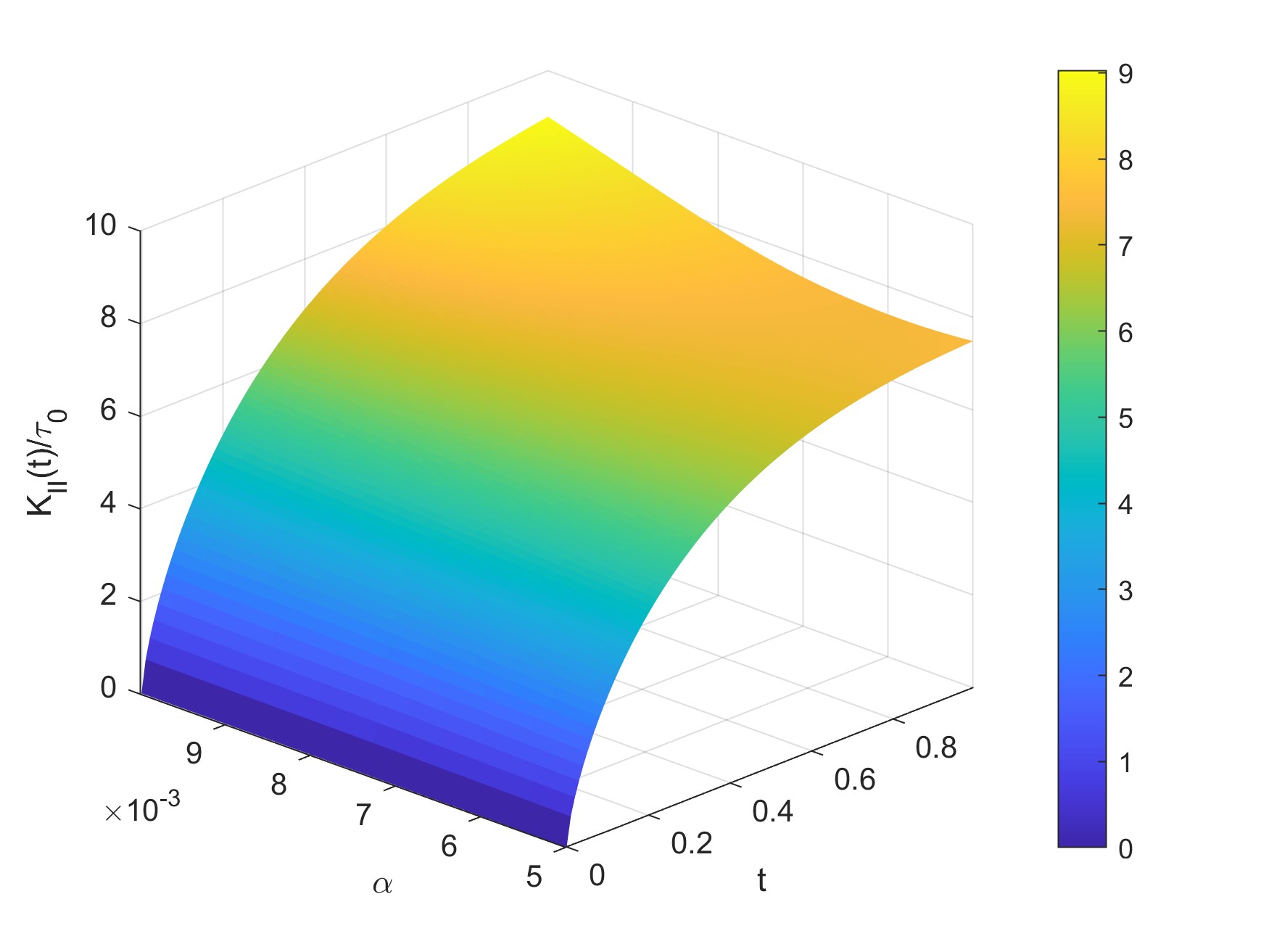}
\caption{}
\label{Fig_12}
\end{subfigure}

\caption{Normalized stress intensity factor ($K_{II}/\tau_0$) of a semi-infinite crack versus time, illustrating the influence of heterogeneity parameter ($\alpha$).}
\label{Figure 6}
\efg

\bfg[htbp]
\centering
\begin{subfigure}[b] {0.8\textwidth}
\includegraphics[width=\textwidth ]{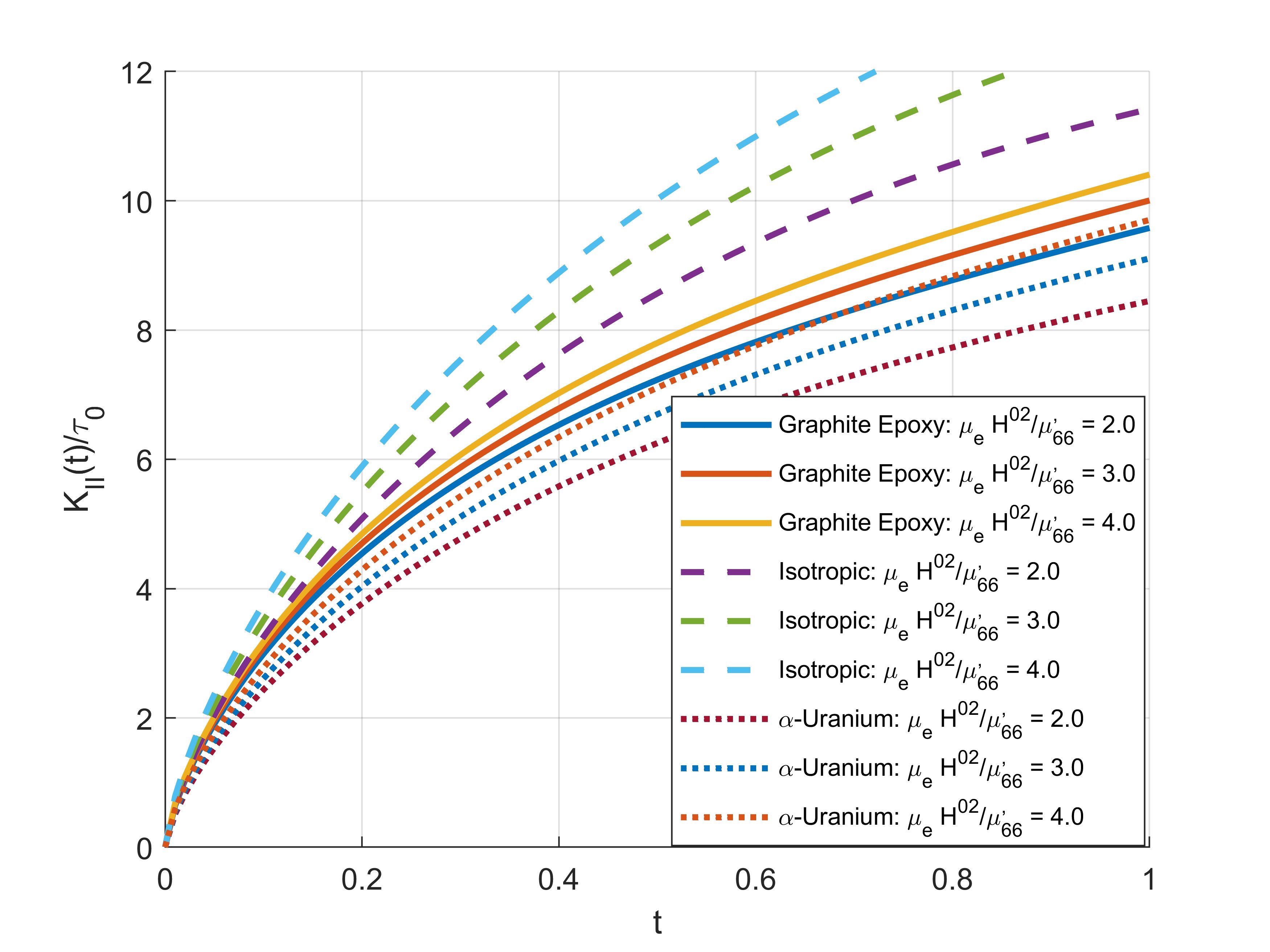}
\caption{}
\label{Fig_13}
\end{subfigure}
~
\begin{subfigure}[b] {0.8\textwidth}
\includegraphics[width=\textwidth ]{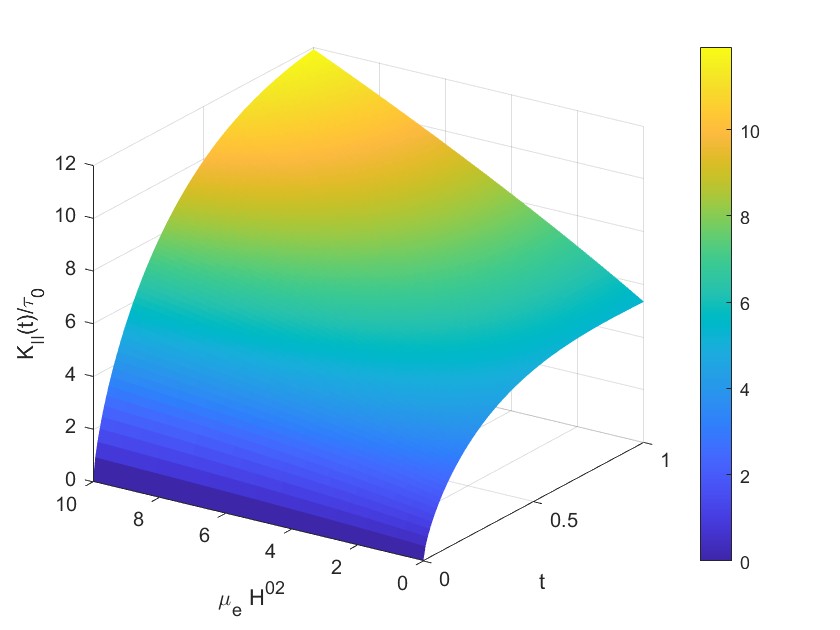}
\caption{}
\label{Fig_14}
\end{subfigure}

\caption{Normalized stress intensity factor ($K_{II}/\tau_0$) of a semi-infinite crack versus time, illustrating the influence of magneto-elastic coupling parameter ($\mu_e {H^0}^2/\mu_{66}'$).}
\label{Figure 7}
\efg

\bfg[htbp]
\centering
\begin{subfigure}[b] {0.8\textwidth}
\includegraphics[width=\textwidth ]{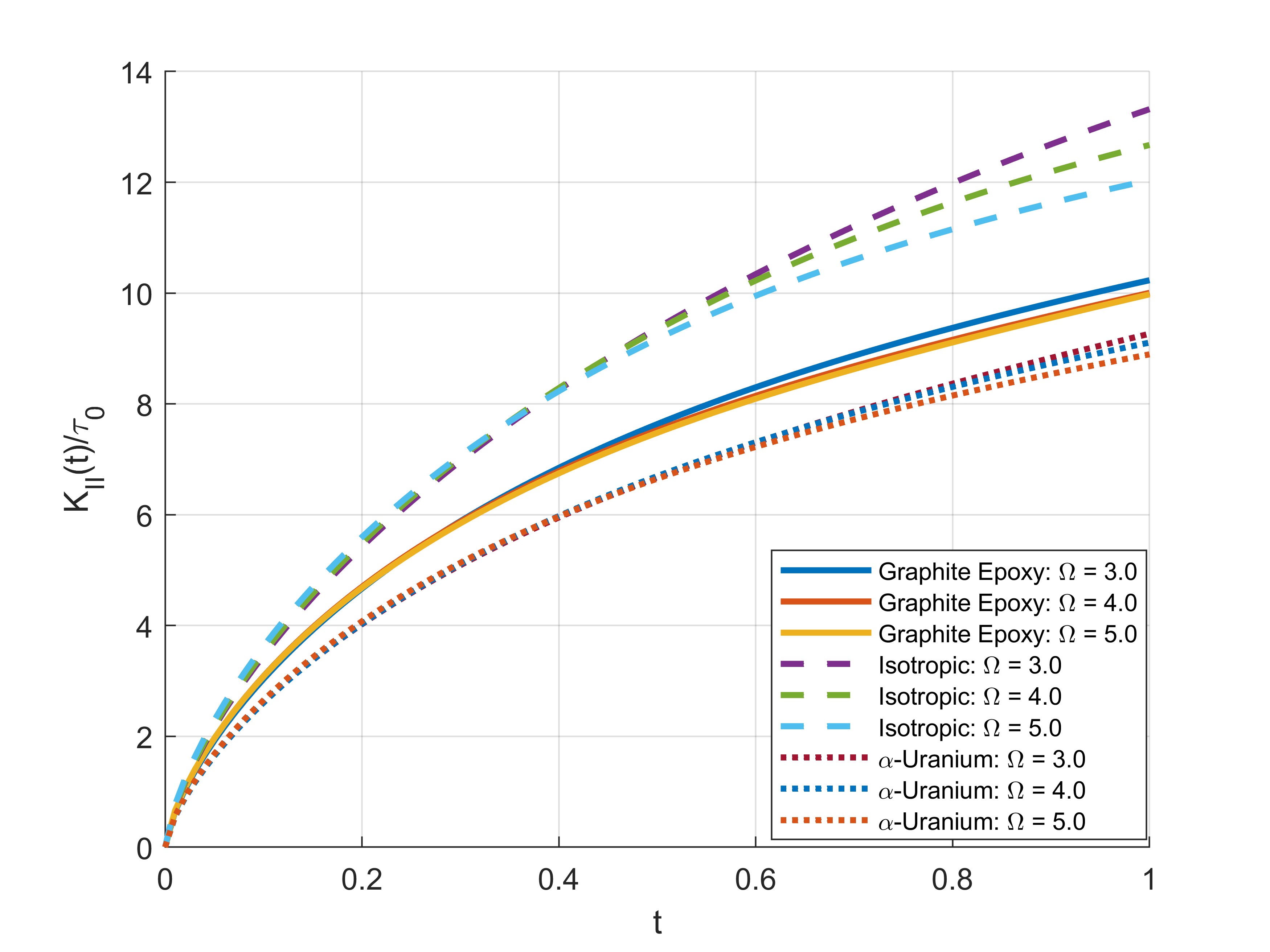}
\caption{}
\label{Fig_15}
\end{subfigure}
~
\begin{subfigure}[b] {0.8\textwidth}
\includegraphics[width=\textwidth ]{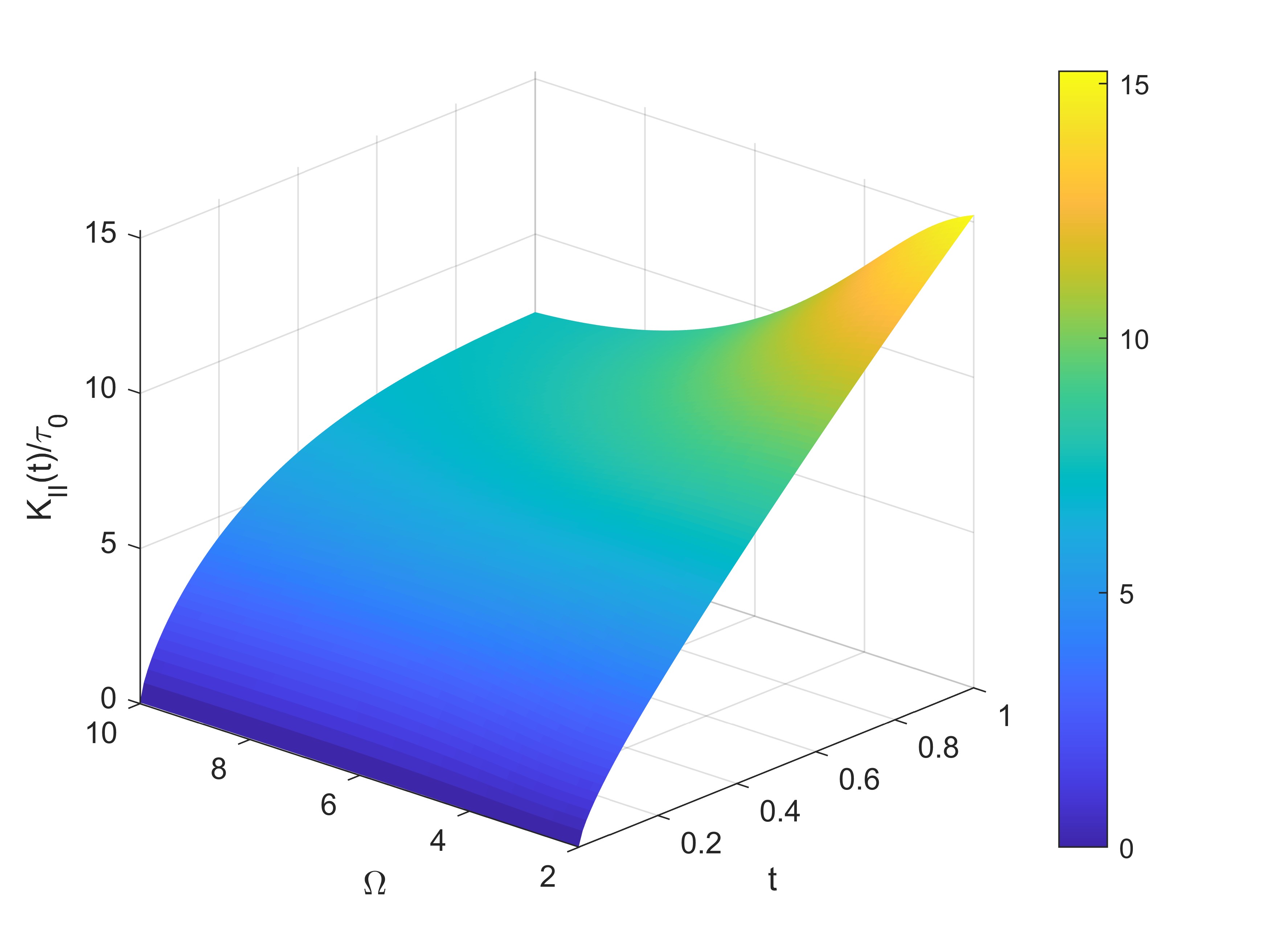}
\caption{}
\label{Fig_16}
\end{subfigure}

\caption{Normalized stress intensity factor ($K_{II}/\tau_0$) of a semi-infinite crack versus time, illustrating the influence of rotational parameter ($\Omega$).}
\label{Figure 8}
\efg

\bfg[htbp]
\centering
\begin{subfigure}[b] {0.8\textwidth}
\includegraphics[width=\textwidth ]{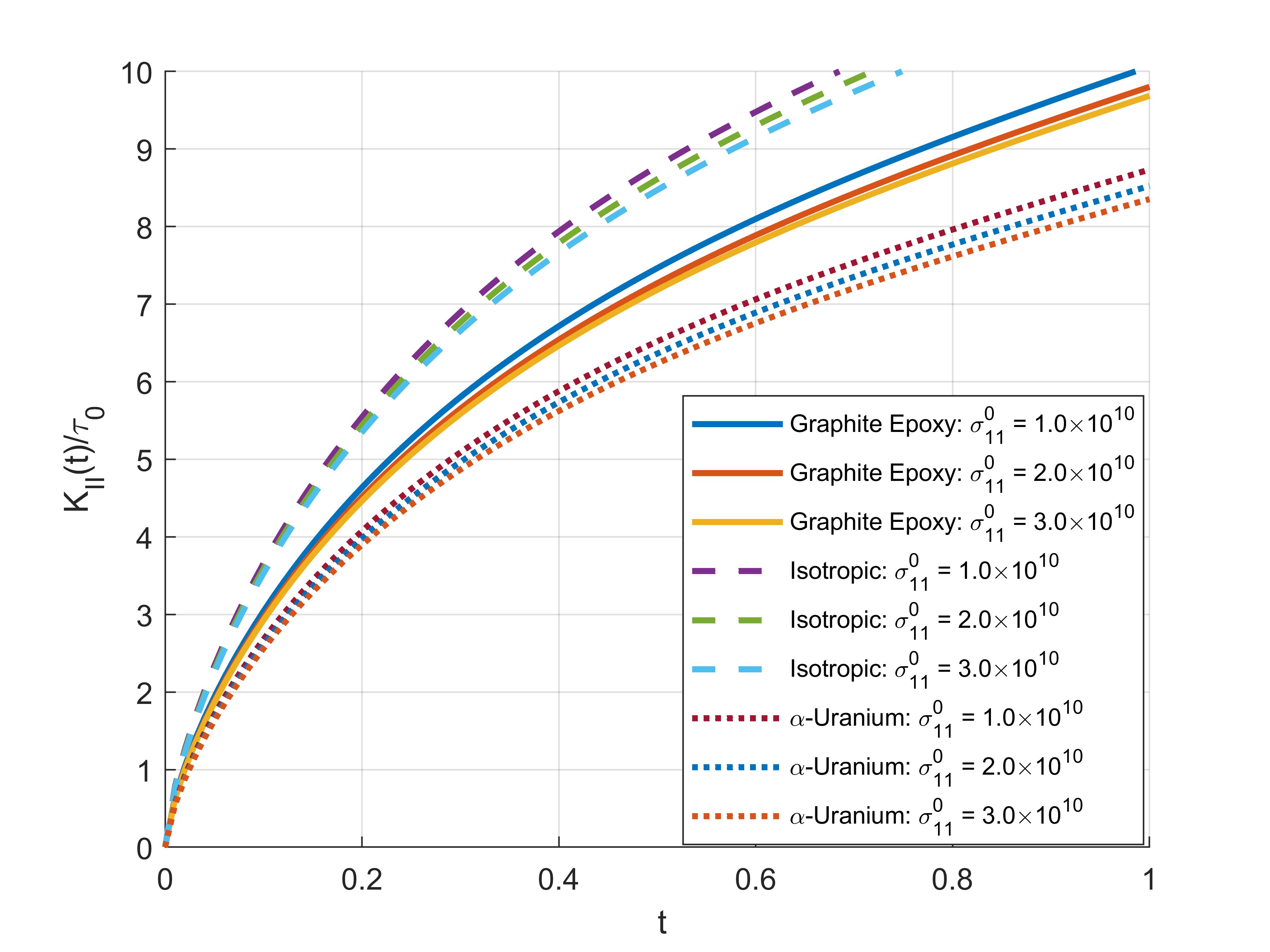}
\caption{}
\label{Fig_17}
\end{subfigure}
~
\begin{subfigure}[b] {0.8\textwidth}
\includegraphics[width=\textwidth ]{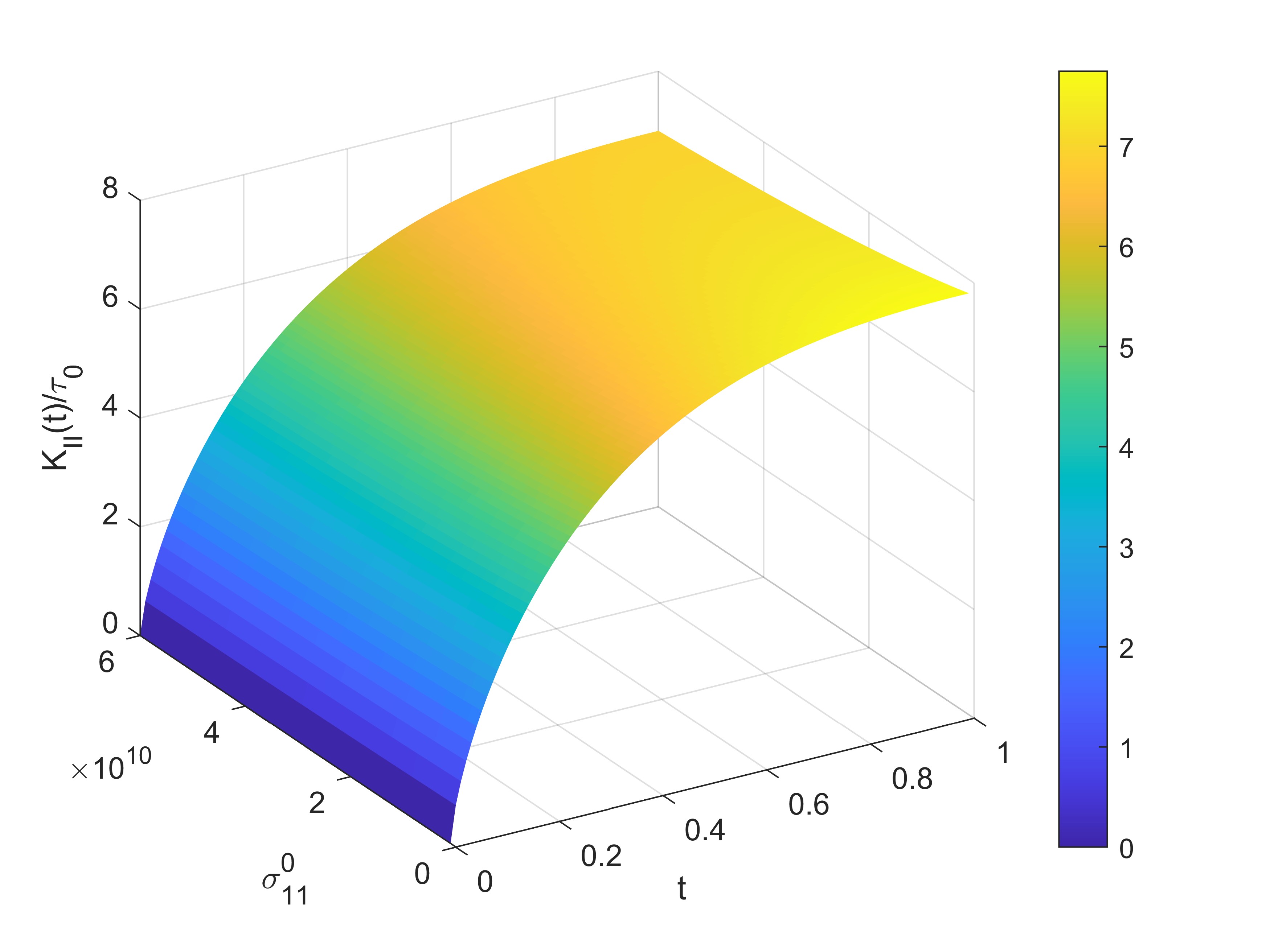}
\caption{}
\label{Fig_18}
\end{subfigure}

\caption{Normalized stress intensity factor ($K_{II}/\tau_0$) of a semi-infinite crack versus time, illustrating the influence of initial stress along the $x_1$-axis ($\sigma_{11}^0$).}
\label{Figure 9}
\efg
\bfg[htbp]
\centering
\begin{subfigure}[b] {0.8\textwidth}
\includegraphics[width=\textwidth ]{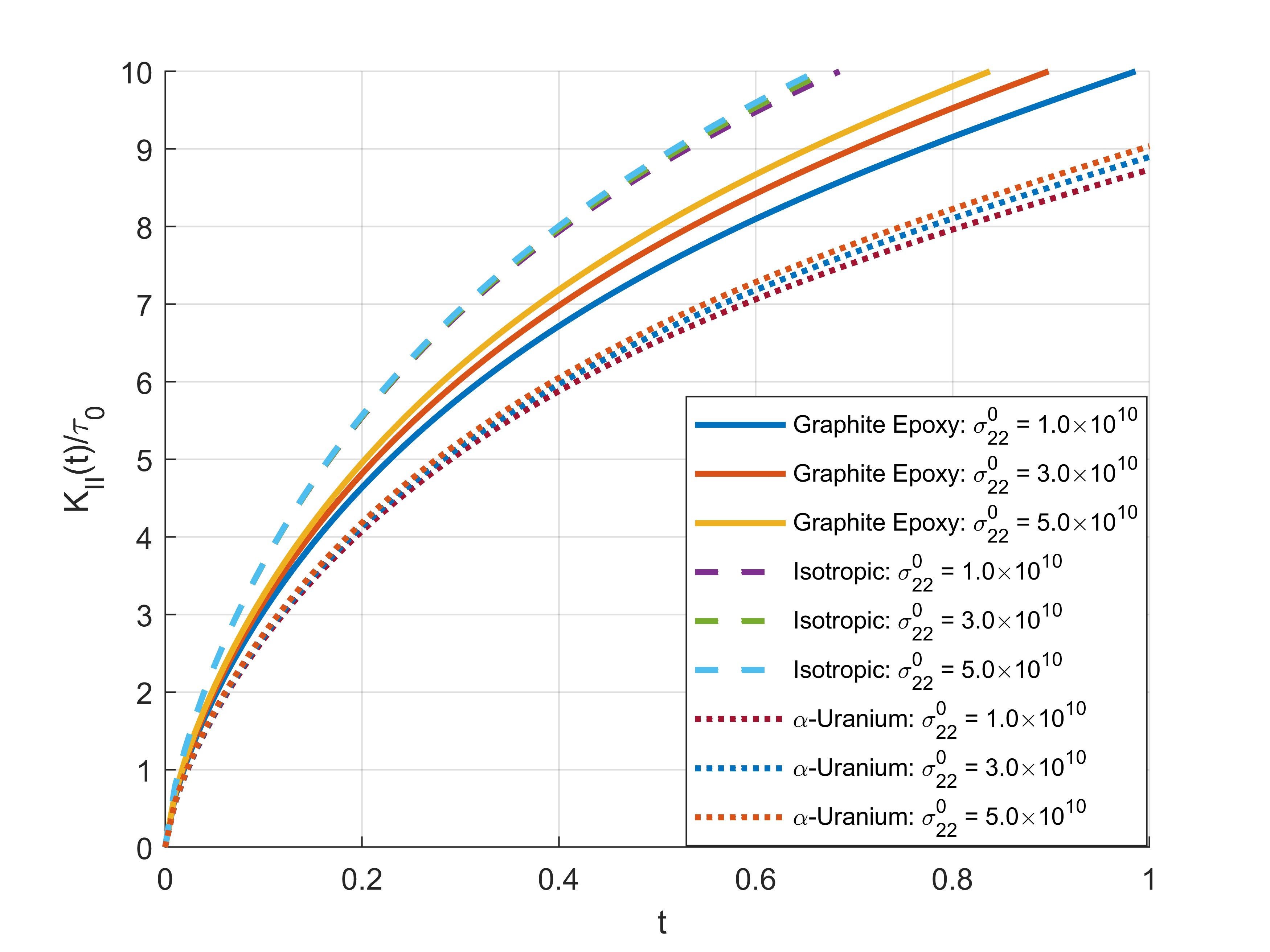}
\caption{}
\label{Fig_19}
\end{subfigure}
~
\begin{subfigure}[b] {0.8\textwidth}
\includegraphics[width=\textwidth ]{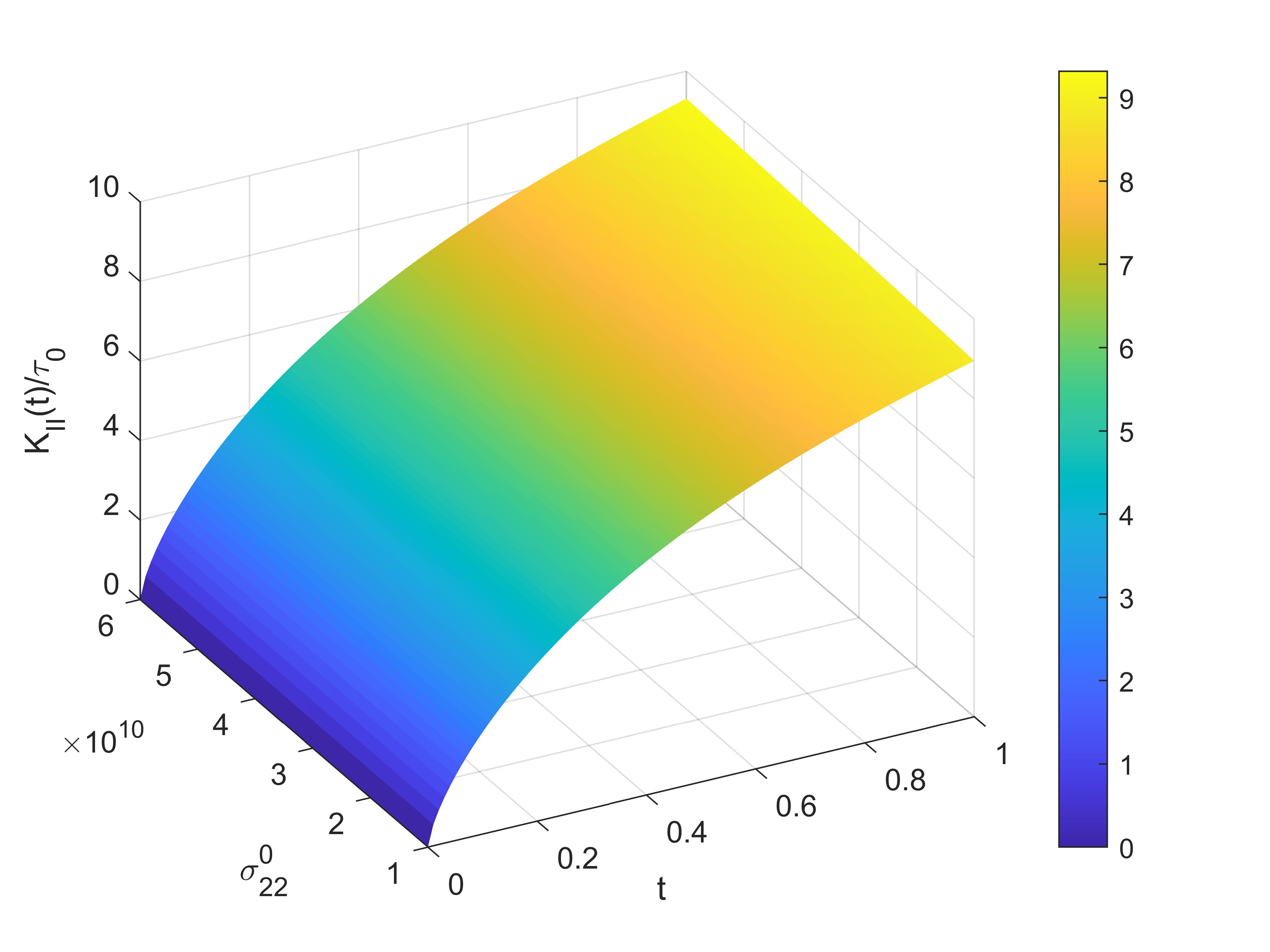}
\caption{}
\label{Fig_20}
\end{subfigure}

\caption{Normalized stress intensity factor ($K_{II}/\tau_0$) of a semi-infinite crack versus time, illustrating the influence of initial stress along the $x_2$-axis ($\sigma_{22}^0$).}
\label{Figure 10}
\efg

\section{Conclusion}
\label{Conclusion}
The response of a semi-infinite crack in a functionally graded magnetoelastic orthotropic strip under longitudinal-axis rotation arises from the combined effects of material gradation, magnetoelastic coupling, and rotational dynamics. Both Mode I (opening) and Mode II (sliding) fractures are analyzed under sudden traction along the crack surfaces, capturing the dynamic response of the system. To this end, the Wiener–Hopf method is employed to develop an analytical formulation, yielding explicit Laplace-domain expressions for the stress intensity factors (SIFs). This formulation allows a systematic investigation of how material heterogeneity, magnetic interactions, and rotation jointly influence crack propagation. The results provide practical insight into fracture behavior in functionally graded magnetoelastic structures and establish a framework for predicting SIFs under dynamic loading conditions. The key contributions of the study are summarized as follows:
\begin{enumerate}
    \item Increasing the heterogeneity of the medium amplifies the time-dependent variation of the stress intensity factor for both Mode I and Mode II, as the gradients in material properties modify the stress distribution along the crack front.  
This effect can be leveraged in designing functionally graded coatings for aerospace and turbine components to optimize fracture resistance throughout their service life.
    \item Increasing the magnetoelasticity of the medium results in a higher SIF for Mode II, due to additional shear coupling between magnetic and elastic fields, while reducing the SIF for Mode I by alleviating normal stress concentration.  
    Such control over mode-specific fracture response is beneficial in magnetoelastic SAW sensors and MRI-compatible actuators, where selective damping of certain wave modes is desired.  

    \item The rotational parameter reduces the stress concentration over time for both loading modes, as rotational inertia introduces a stabilizing effect that delays crack-tip stress accumulation.  
    This effect is valuable for high-speed rotating machinery, maglev train components, and energy-storage flywheels, where dynamic fracture suppression is critical.  
    \item Increasing the magnitude of horizontal initial stress results in an increased SIF for Mode I, as tensile stresses perpendicular to the crack plane amplify crack opening, and a decreased SIF for Mode II by reducing in-plane shear concentration.  
This effect is significant in underground tunnel linings and long-span bridges, where horizontal ground stress can dominate fracture behavior.

\item Increasing vertical initial stress leads to a decreased SIF in Mode I, as compressive stresses normal to the crack plane inhibit crack opening, and an increased SIF in Mode II due to enhanced shear stress along the crack front.  
This mechanism is relevant for deep geological repositories, high-rise building foundations, and layered rock masses, where vertical overburden stress influences shear-driven failure.

\end{enumerate}
The findings of this study advance the understanding of crack behavior across diverse scenarios and provide a rigorous framework for the analytical modeling of semi-infinite cracks in heterogeneous magnetoelastic orthotropic materials.

\noindent {\bf Acknowledgements}\\
The authors express their sincere gratitude to the National Institute of Technology Hamirpur for providing the research facilities to Ms. Diksha during her doctoral studies. The financial assistance received from the University Grants Commission (UGC) in the form of a fellowship is gratefully acknowledged.\\
\noindent {\bf Conflicts of interest}\\
The authors declare no conflict of interest.\\
\noindent {\bf Code availability}\\
All computational codes employed in this study were independently developed by the authors in MATLAB, without reliance on third-party software or external code libraries.

\bibliographystyle{unsrt}
\bibliography{refrences}

\begin{thebibliography}{10}

\bibitem{erdogan1995fracture}
Fazil Erdogan.
\newblock Fracture mechanics of functionally graded materials.
\newblock {\em Composites Engineering}, 5(7):753--770, 1995.

\bibitem{gu1997cracks}
Pei Gu and RJ~Asaro.
\newblock Cracks in functionally graded materials.
\newblock {\em International Journal of solids and structures}, 34(1):1--17, 1997.

\bibitem{bahr2003cracks}
H-A Bahr, H~Balke, T~Fett, I~Hofinger, G~Kirchhoff, D~Munz, A~Neubrand, AS~Semenov, H-J Weiss, and YY~Yang.
\newblock Cracks in functionally graded materials.
\newblock {\em Materials Science and Engineering: A}, 362(1-2):2--16, 2003.

\bibitem{yu2009identification}
Zhigang Yu and Fulei Chu.
\newblock Identification of crack in functionally graded material beams using the p-version of finite element method.
\newblock {\em Journal of Sound and Vibration}, 325(1-2):69--84, 2009.

\bibitem{zhang20113d}
CH~Zhang, M~Cui, J~Wang, XW~Gao, J~Sladek, and V~Sladek.
\newblock 3d crack analysis in functionally graded materials.
\newblock {\em Engineering Fracture Mechanics}, 78(3):585--604, 2011.

\bibitem{sadowski2009cracks}
T~Sadowski, L~Marsavina, N~Peride, and E-M Craciun.
\newblock Cracks propagation and interaction in an orthotropic elastic material: Analytical and numerical methods.
\newblock {\em Computational Materials Science}, 46(3):687--693, 2009.

\bibitem{singh2025review}
Ritika Singh.
\newblock A review on edge crack problems in orthotropic material.
\newblock {\em Mathematics and Mechanics of Solids}, page 10812865251342229, 2025.

\bibitem{gay2022composite}
Daniel Gay.
\newblock {\em Composite materials: design and applications}.
\newblock CRC press, 2022.

\bibitem{parveez2022scientific}
Bisma Parveez, MI~Kittur, Irfan~Anjum Badruddin, Sarfaraz Kamangar, Mohamed Hussien, and MA~Umarfarooq.
\newblock Scientific advancements in composite materials for aircraft applications: a review.
\newblock {\em Polymers}, 14(22):5007, 2022.

\bibitem{aspri2022dislocations}
Andrea Aspri, Elena Beretta, and Anna Mazzucato.
\newblock Dislocations in a layered elastic medium with applications to fault detection.
\newblock {\em Journal of the European Mathematical Society}, 25(3):1091--1112, 2022.

\bibitem{reches2023earthquakes}
Ze'ev Reches and Jay Fineberg.
\newblock Earthquakes as dynamic fracture phenomena.
\newblock {\em Journal of Geophysical Research: Solid Earth}, 128(3):e2022JB026295, 2023.

\bibitem{rubio2000dynamic}
C~Rubio-Gonzalez and JJ~Mason.
\newblock Dynamic stress intensity factors at the tip of a uniformly loaded semi-infinite crack in an orthotropic material.
\newblock {\em Journal of the Mechanics and Physics of Solids}, 48(5):899--925, 2000.

\bibitem{karan2024interaction}
Somashri Karan, Palas Mandal, Sanjoy Basu, and Subhas~Chandra Mandal.
\newblock Interaction of shear waves with semi-infinite moving crack inside of a orthotropic media.
\newblock {\em Waves in Random and Complex Media}, 34(5):4010--4026, 2024.

\bibitem{zhong2012thermoelastic}
Xian-Ci Zhong and Bing Wu.
\newblock Thermoelastic analysis for an opening crack in an orthotropic material.
\newblock {\em International journal of fracture}, 173(1):49--55, 2012.

\bibitem{ayhan2006fracture}
AO~Ayhan, AC~Kaya, A~Loghin, JH~Laflen, RD~McClain, and D~Slavik.
\newblock Fracture analysis of cracks in orthotropic materials using ansys{\textregistered}.
\newblock In {\em Turbo Expo: Power for Land, Sea, and Air}, volume 42401, pages 873--881, 2006.

\bibitem{fakoor2017influence}
M~Fakoor, R~Rafiee, and M~Sheikhansari.
\newblock The influence of fiber-crack angle on the crack tip parameters in orthotropic materials.
\newblock {\em Proceedings of the Institution of Mechanical Engineers, Part C: Journal of Mechanical Engineering Science}, 231(3):418--431, 2017.

\bibitem{farid2019mixed}
Hannaneh~Manafi Farid and Mahdi Fakoor.
\newblock Mixed mode i/ii fracture criterion for arbitrary cracks in orthotropic materials considering t-stress effects.
\newblock {\em Theoretical and Applied Fracture Mechanics}, 99:147--160, 2019.

\bibitem{xu2008dynamic}
Hongmin Xu, Xuefeng Yao, Xiqiao Feng, and Yang~Yeh Hisen.
\newblock Dynamic stress intensity factors of a semi-infinite crack in an orthotropic functionally graded material.
\newblock {\em Mechanics of Materials}, 40(1-2):37--47, 2008.

\bibitem{ustinov2020orthotropic}
Konstantin~B Ustinov, Roberta Massabo, and Dmitry~S Lisovenko.
\newblock Orthotropic strip with central semi-infinite crack under arbitrary loads applied far apart from the crack tip. analytical solution.
\newblock {\em Engineering Failure Analysis}, 110:104410, 2020.

\bibitem{knauss1966stresses}
Wolfgang~Gustav Knauss.
\newblock Stresses in an infinite strip containing a semi-infinite crack.
\newblock {\em Journal of Applied Mechanics}, 33(2):356--362, 1966.

\bibitem{das2024study}
Subir Das and Anshika Tanwar.
\newblock Study of an interfacial semi-infinite crack in a composite structure.
\newblock {\em Acta Mechanica}, 235(8):4961--4977, 2024.

\bibitem{ustinov2019semi}
Konstantin Ustinov.
\newblock On semi-infinite interface crack in bi-material elastic layer.
\newblock {\em European Journal of Mechanics-A/Solids}, 75:56--69, 2019.

\bibitem{chen2014propagation}
Hao-sen Chen, Wei-yi Wei, Jin-xi Liu, and Dai-ning Fang.
\newblock Propagation of a semi-infinite conducting crack in piezoelectric materials: Mode-i problem.
\newblock {\em Journal of the Mechanics and Physics of Solids}, 68:77--92, 2014.

\bibitem{krylov1977handbook}
Vladimir~Ivanovich Krylov, Nadezhda~Sergeevna Skoblia, and George Yankovsky.
\newblock A handbook of methods of approximate fourier transformation and inversion of the laplace transformation.
\newblock {\em (No Title)}, 1977.

\bibitem{singh2020semi}
A~Singh, S~Das, H~Altenbach, and E-M Craciun.
\newblock Semi-infinite moving crack in an orthotropic strip sandwiched between two identical half planes.
\newblock {\em ZAMM-Journal of Applied Mathematics and Mechanics/Zeitschrift f{\"u}r Angewandte Mathematik und Mechanik}, 100(2):e201900202, 2020.

\bibitem{das2024wiener}
Shiv~Shankar Das, Anshika Tanwar, Subir Das, and Eduard-Marius Craciun.
\newblock Wiener--hopf method to solve the anti-plane problem of moving semi-infinite crack in orthotropic composite materials.
\newblock {\em Mathematics and Mechanics of Solids}, 29(7):1311--1324, 2024.

\bibitem{alam2024magnetoelastic}
Samim Alam, Somashri Karan, and Subhas~Chandra Mandal.
\newblock Magnetoelastic effect on semi-infinite moving crack in an orthotropic structure influenced by anti-plane wave.
\newblock {\em ZAMM-Journal of Applied Mathematics and Mechanics/Zeitschrift f{\"u}r Angewandte Mathematik und Mechanik}, 104(5):e202300825, 2024.

\bibitem{ing2001transient}
Yi-Shyong Ing and Chien-Ching Ma.
\newblock Transient response of a surface crack subjected to dynamic anti-plane concentrated loadings.
\newblock {\em International journal of fracture}, 109(3):239--261, 2001.

\bibitem{flitman1963waves}
LM~Flitman.
\newblock Waves caused by a sudden crack in a continuous elastic medium.
\newblock {\em Journal of Applied Mathematics and Mechanics}, 27(4):938--953, 1963.

\bibitem{khimin2024analysis}
Denis Khimin, Johannes Lankeit, Marc~C Steinbach, and Thomas Wick.
\newblock Analysis of a space-time phase-field fracture complementarity model and its optimal control formulation.
\newblock {\em SIAM Journal on Mathematical Analysis}, 56(5):6192--6212, 2024.

\bibitem{bresciani2025quasistatic}
Marco Bresciani and Manuel Friedrich.
\newblock Quasistatic growth of cavities and cracks in the plane.
\newblock {\em SIAM Journal on Mathematical Analysis}, 57(3):2287--2315, 2025.

\bibitem{abrate1991impact}
Serge Abrate.
\newblock Impact on laminated composite materials.
\newblock 1991.

\bibitem{shi2017modelling}
Yu~Shi and Constantinos Soutis.
\newblock Modelling low velocity impact induced damage in composite laminates.
\newblock {\em Mechanics of Advanced Materials and Modern Processes}, 3(1):14, 2017.

\bibitem{suresh1998fatigue}
Subra Suresh.
\newblock {\em Fatigue of materials}.
\newblock Cambridge university press, 1998.

\bibitem{he2025damage}
Zhenhong He, Xiaoqi Chen, Xiaoqiang Zhang, Yongbo Jiang, Xianben Ren, and Ying Li.
\newblock Damage prediction of hull structure under near-field underwater explosion based on machine learning.
\newblock {\em Applied Ocean Research}, 154:104329, 2025.

\bibitem{wang2001dynamic}
CY~Wang, C~Rubio-Gonzalez, and JJ~Mason.
\newblock The dynamic stress intensity factor for a semi-infinite crack in orthotropic materials with concentrated shear impact loads.
\newblock {\em International journal of solids and structures}, 38(8):1265--1280, 2001.

\bibitem{alam2025mode}
Samim Alam and Subhas Chandra~Mandal.
\newblock Mode-i moving semi-infinite crack in an infinitely long orthotropic strip in the presence of electromagnetic field.
\newblock {\em Fatigue \& Fracture of Engineering Materials \& Structures}, 48(4):1479--1495, 2025.

\bibitem{deshpande2018effect}
Revati~Rajeev Deshpande.
\newblock {\em Effect of Centrifugal Stiffening on the Natural Frequencies of Aircraft Wings During Rapid Roll Maneuvers}.
\newblock PhD thesis, Virginia Tech, 2018.

\bibitem{luo2024magnetoelectric}
Bin Luo, Prasanth Velvaluri, Yisi Liu, and Nian-Xiang Sun.
\newblock Magnetoelectric baw and saw devices: a review.
\newblock {\em Micromachines}, 15(12):1471, 2024.

\bibitem{wolframm2022high}
Henrik Wolframm, Viktor Schell, Eckhard Quandt, Michael H{\"o}ft, and Andreas Bahr.
\newblock High sensitivity magnetoelastic surface acoustic wave (saw) magnetic field sensors.
\newblock In {\em Spintronics XV}, volume 12205, pages 25--31. SPIE, 2022.

\bibitem{jweeg2020dynamic}
Muhsin~J Jweeg, Salah~N Alnomani, and Salah~K Mohammad.
\newblock Dynamic analysis of a rotating stepped shaft with and without defects.
\newblock In {\em IOP Conference Series: Materials Science and Engineering}, volume 671, page 012004. IOP Publishing, 2020.

\bibitem{valverde2022influence}
Borja Valverde-Marcos, B~Mu{\~n}oz-Abella, P~Rubio, and L~Rubio.
\newblock Influence of the rotation speed on the dynamic behaviour of a cracked rotating beam.
\newblock {\em Theoretical and applied fracture mechanics}, 117:103209, 2022.

\bibitem{quaranta2014rotorcraft}
Giuseppe Quaranta, Aykut Tamer, Vincenzo Muscarello, Pierangelo Masarati, Massimo Gennaretti, Jacopo Serafini, and Marco~Molica Colella.
\newblock Rotorcraft aeroelastic stability using robust analysis.
\newblock {\em CEAS Aeronautical Journal}, 5(1):29--39, 2014.

\bibitem{senturk2016design}
Yusuf~Mert Senturk and Volkan Patoglu.
\newblock Design and control of an mri compatible series elastic actuator.
\newblock In {\em 2016 IEEE international conference on robotics and biomimetics (ROBIO)}, pages 1473--1479. IEEE, 2016.

\bibitem{chen2007wave}
Jiangyi Chen, E~Pan, and Hualing Chen.
\newblock Wave propagation in magneto-electro-elastic multilayered plates.
\newblock {\em International journal of Solids and Structures}, 44(3-4):1073--1085, 2007.

\bibitem{xiang2022dynamic}
Huoyue Xiang, Xiangfu Tian, Yongle Li, and Min Zeng.
\newblock Dynamic interaction analysis of high-speed maglev train and guideway with a control loop failure.
\newblock {\em International Journal of Structural Stability and Dynamics}, 22(10):2241012, 2022.

\bibitem{shi2003indentation}
Dongai Shi, Yuan Lin, and Timothy~C Ovaert.
\newblock Indentation of an orthotropic half-space by a rigid ellipsoidal indenter.
\newblock {\em J. Trib.}, 125(2):223--231, 2003.

\bibitem{biot1940influence}
Maurice~A Biot.
\newblock The influence of initial stress on elastic waves.
\newblock {\em Journal of Applied Physics}, 11(8):522--530, 1940.

\bibitem{chaudhary2025integral}
Soniya Chaudhary, Pawan~Kumar Sharma, Qasem~M Al-Mdallal, et~al.
\newblock Integral transform technique for determining stress intensity factor in wave propagation through functionally graded piezoelectric-viscoelastic structure.
\newblock {\em Computers \& Mathematics with Applications}, 186:130--154, 2025.

\bibitem{chaudhary2025crack}
Soniya Chaudhary, Diksha, and Pawan~Kumar Sharma.
\newblock Crack dynamics in rotating, initially stressed material strip: A mathematical approach.
\newblock {\em Applied Mathematical Modelling}, 140:115916, 2025.

\bibitem{noble1959methods}
Ben Noble and George Weiss.
\newblock Methods based on the wiener-hopf technique for the solution of partial differential equations.
\newblock {\em Physics Today}, 12(9):50--50, 1959.

\bibitem{freund1998dynamic}
Lambert~Ben Freund and LB~Freud.
\newblock {\em Dynamic fracture mechanics}.
\newblock Cambridge university press, 1998.

\end{thebibliography}
\end{document}